\documentclass[aps,prl,reprint,superscriptaddress]{revtex4-1}

\usepackage{graphicx}% Include figure files
\usepackage{dcolumn}% Align table columns on decimal point
\usepackage{bm}% bold math
\usepackage{braket}
\usepackage{amsmath,MnSymbol,nicefrac}
\usepackage{subcaption}
\usepackage{cleveref}

\newcommand{\citekohn}{Kohn_Variational_1948,Nesbet_Analysis_1968,Nesbet_Anomaly_1969,Rescigno_Electron_1979,McCurdy_Interrelation_1987,Zhang_Quantum_1988,Rescigno_Complex_1995}

\newcommand{\citekohnimp}{Rescigno_Separable_1981,Rescigno_Disappearance_1988,Schneider_Complex_1988,Orel_Variational_1990,Rescigno_Continuum_1991,McCurdy_Beyond_1992,Rescigno_Algebraic_1997,Rescigno_Improvements_1998,McCurdy_Calculation_1998}

\newcommand{\citekohnphoto}{Rescigno_Interchannel_1993,Orel_Photoionization_1997,Miyabe_Theoretical_2009,Sann_Electron_2011,Williams_Imaging_2012,Douguet_Time_2012,Marggi_Poullain_Recoil_2014,Fonseca_dos_Santos_Ligand_2015,McCurdy_Unambiguous_2017,Champenois_Ultrafast_2019}

\newcommand{\citekohnelec}{Hazi_AbInitio_1981,Rescigno_Accurate_1989,Schneider_Accurate_1991,Lengsfield_Electron_1991,Gil_Polarization_1993,Rescigno_Theoretical_1999,Rescigno_Dynamics_2006,Adaniya_Imaging_2009,Slaughter_Ion_2016,Rescigno_Dynamics_2016}

\begin{document}

\title{Electron Correlation Effects in Attosecond Photoionization of CO$_2$}% Force line breaks with \\

\author{Andrei~Kamalov}
\email{andrei.kamalov@gmail.com}
\affiliation{Stanford PULSE Institute, SLAC National Accelerator Laboratory, Menlo Park, CA, USA}
\affiliation{Department of Physics, Stanford University, Stanford, CA, USA}

\author{Anna~L.~Wang}
\affiliation{Stanford PULSE Institute, SLAC National Accelerator Laboratory, Menlo Park, CA, USA}
\affiliation{Department of Applied Physics, Stanford University, Stanford, CA, USA}

\author{Philip~H.~Bucksbaum}
\affiliation{Stanford PULSE Institute, SLAC National Accelerator Laboratory, Menlo Park, CA, USA}
\affiliation{Department of Physics, Stanford University, Stanford, CA, USA}
\affiliation{Department of Applied Physics, Stanford University, Stanford, CA, USA}

\author{Daniel~J.~Haxton}
\affiliation{KLA Corporation, Milpitas, CA, USA}

\author{James~P.~Cryan}
\email{jcryan@slac.stanford.edu}
\affiliation{Stanford PULSE Institute, SLAC National Accelerator Laboratory, Menlo Park, CA, USA}
\affiliation{Linac Coherent Light Source, SLAC National Accelerator Laboratory, Menlo Park, CA, USA}

\date{\today}

\begin{abstract}
A technique for measuring photoionization time delays with attosecond precision is combined with calculations of photoionization matrix elements to demonstrate how multi-electron dynamics affect photoionization time delays in carbon dioxide.
Electron correlation is observed to affect the time delays through two mechanisms: autoionization of molecular Rydberg states and accelerated escape from a continuum shape resonance.
\end{abstract}

\maketitle

%%%%%% Introduction %%%%%%%%%%%%%%%
% Photoionization is a fundamental scattering process, in which light energy is converted to chemical degrees of freedom.  
% The quantum scattering process 
%Photoionization is a basic quantum scattering process involving the rearrangement of multiple degrees of freedom.%~(photon$+$neutral molecule to electron$+$cation).  
Photoionization is a basic quantum scattering process involving the rearrangement of degrees freedom in the total system.
In the time domain, this is described by an incoming photon wavepacket that couples to outgoing electron wavepackets~(EWPs) in the final-state cation channels.
The term ``photoionization time delay'' refers to the time required for a photoionized EWP to propagate out of the electric potential of the residual cation.
It may be defined semi-classically as the extra time required to propagate a photoelectron from its birth location to a detector position, compared to some reference~\cite{dahlstrom_introduction_2012,klunder_probing_2011,serov_interpretation_2013-1}. 
Recent advances in the production of attosecond laser pulses have enabled direct probing of these delays~\cite{pazourek_attosecond_2015}.
Combining these time delay measurements with theoretical modeling 
%can provide an invertible map to the 
reveals the underlying
quantum dynamics of the photoionization process~\cite{dahlstrom_introduction_2012,pazourek_attosecond_2015,schultze_delay_2010,klunder_probing_2011,dahlstrom_introduction_2012,guenot_photoemission-time-delay_2012,kheifets_time_2013,serov_interpretation_2013-1,guenot_measurements_2014,palatchi_atomic_2014,sabbar_resonance_2015,ossiander_attosecond_2017,isinger_photoionization_2017,kiesewetter_probing_2017,busto_timefrequency_2018}.

Attosecond electron dynamics of ionization are necessarily violent because additional kinetic energy must be imparted to the bound electron for it to escape the Coulomb potential.
The added kinetic energy may be redistributed through Coulomb and exchange scattering with other electrons, dynamically modifying this ionic potential.
This is particularly important in molecular systems.
Although the asymptotic state of the total system~(cation plus ionized electron) is easily understood in a single-electron picture, this picture may break down when the the electron has not yet escaped into the asymptotic region for detection.
The modification of the ionic potential is imprinted onto the measured photoionization time delays~\cite{pazourek_attosecond_2015}, which provides direct access to the temporal evolution of electron-electron interactions.
Previous measurements of photoionization time delays have made use of this effect, which has led to a deeper understanding of electron correlations in shake-up ionization~\cite{ossiander_attosecond_2017} and atomic autoionization~\cite{gruson_attosecond_2016, kotur_spectral_2016,cirelli_anisotropic_2018,busto_timefrequency_2018}.
The present work combines measurements of the photoionization time delays with numerical calculations of photoionization probability amplitudes to demonstrate how multi-electron dynamics affect ionized EWPs in a molecular system.

%Multi-electron dynamics leaves a clear signature in the measured time delays of the EWPs near autoionizing resonances and molecular shape resonances.
These dynamics leave a clear signature in the measured photoionization time delay in the vicinity of autoionizing and molecular shape resonances.
The enhancement of electron correlation effects near molecular shape resonances was previously considered by Siggel~\textit{et al.}.
They found that the photoelectron angular distribution can be sensitive to multi-electron channel coupling phenomena~\cite{siggel_shaperesonanceenhanced_1993}.
The scattering angle is one of two semi-classical scattering observables; the other is time delay.
The effect that electron-electron interactions would have on the interpretation of photoionization time delay measurements has yet to be considered in the literature.

\begin{figure*}
%\centering
    \includegraphics[width=0.7\textwidth]{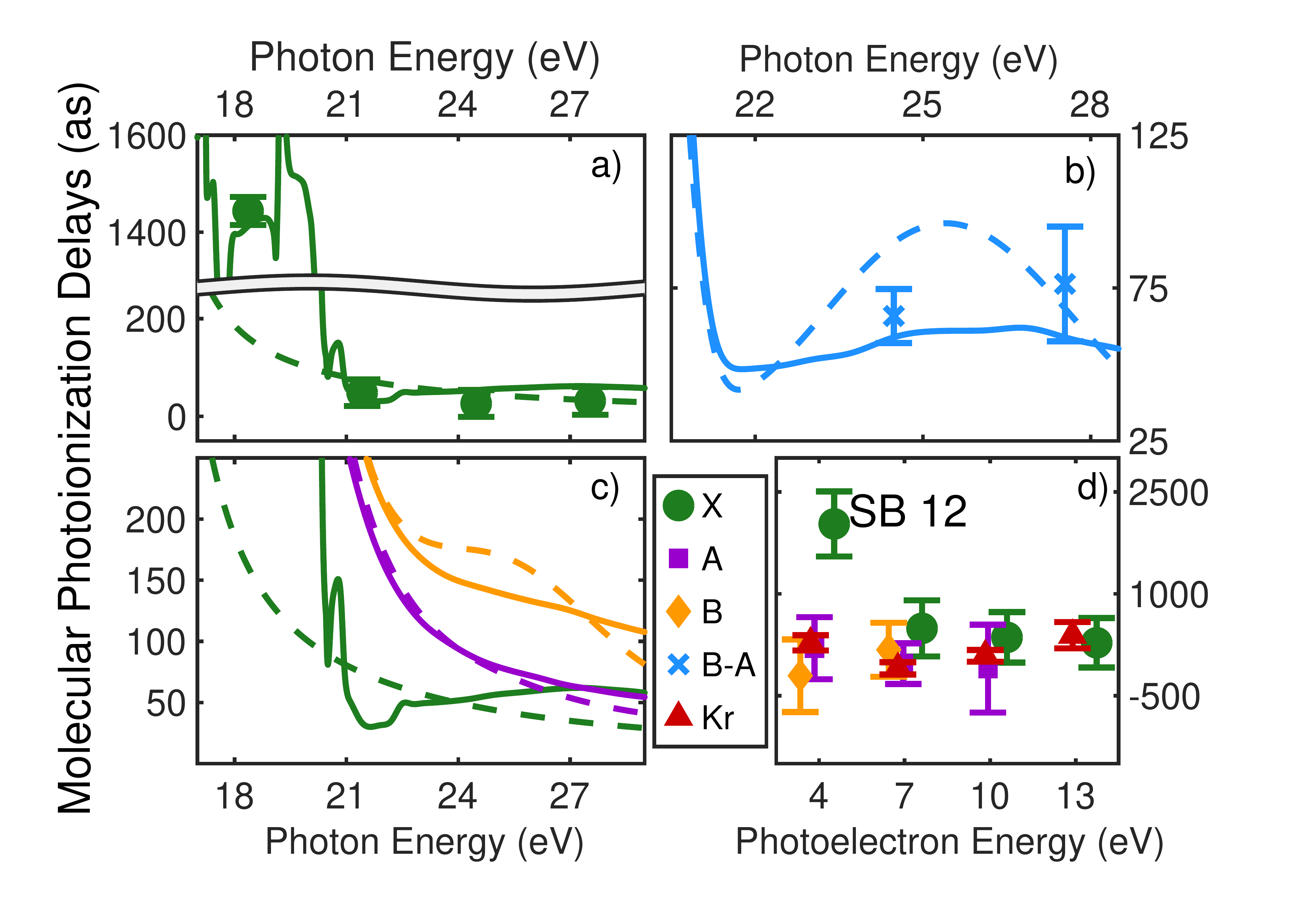}
    \begin{subfigure}{0em}\phantomsubcaption\label{fig::XState}\end{subfigure}
    \begin{subfigure}{0em}\phantomsubcaption\label{fig::ABState}\end{subfigure}
    \begin{subfigure}{0em}\phantomsubcaption\label{fig::Theory}
    \end{subfigure}
    \begin{subfigure}{0em}\phantomsubcaption\label{fig::Rainbow}\end{subfigure}
    \caption{Molecular photoionization time delays for the $X ^{2}\Pi_{g}$~(panel~a), $A ^{2}\Pi_{u}$, and $B ^{2}\Sigma^{+}_{u}$~(panel~b) states of unaligned CO$^{+}_2$.  Panel~c shows the finite difference approximation to the derivative of the scattering phase calculated using the complex-Kohn method and incoherently summed across all molecular orientations and electron emission angles. Calculations are done in two approximations~(see text), the independent channel approximation~(dashed) and coupled-channel method~(solid). We compare the calculations to the measured data~(panels~a~and~b). The long delay reported for the low energy features in the $X$-state channel are extracted from the measurement of the sideband slope shown in panel~d. In all panels, the errorbars represent the $\pm2\sigma$ confidence level.}
    \label{fig::dataWithTheory}
\end{figure*}

Carbon dioxide~(CO$_2$) provides a particularly striking example of multi-electron dynamics in molecular photoionization~\cite{lucchese_studies_1982,lucchese_effects_1990,Lucchese_comparative_1981,harvey_r-matrix_2014}.
Straightforward close-coupling expansions require ninety-six individual cation configurations to reproduce the experimental cross section~\cite{harvey_r-matrix_2014}, far more states than are energetically available as photoionization channels. 
The virtual excitations of the closed cationic channels correspond to multiple-electron excitations of the molecular system, which affect the magnitude and phase of the EWPs escaping into the energetically open channels.
In the present work we measure the photoionization time delays for CO$_2$ and demonstrate that agreement with calculated time delays is contingent upon including electron correlation effects in the calculation.

%%%%%% End Introduction %%%%%%%%%%%%%%%

%%%%%%% Begin Discussion of Data %%%%%%%%%

%Photoionization time delays are extracted from photoelectron spectrograms produced by the combination of a train of attosecond pulses from an extreme ultraviolet~(XUV) frequency comb temporally overlapped with a weak dressing laser-field at half the frequency of the comb spacing.
%The resulting delay-dependent interference patterns reveal the measured time delays for the $X^{2}\Pi_{g}$, $A^{2}\Pi_{u}$, and $B^{2}\Sigma^{+}_{u}$ cationic channels shown in \Cref{fig::dataWithTheory}.
\Cref{fig::dataWithTheory} shows the measured photoionization time delays for the $X^{2}\Pi_{g}$, $A^{2}\Pi_{u}$, and $B^{2}\Sigma^{+}_{u}$ cationic states of CO$_2$.
Details of the measurement procedure are given after a discussion of the results. 
In order to understand the dynamics captured in the time delay measurements, we compare these data with predicted delays calculated using an implementation~\cite{\citekohnimp} of the complex Kohn variational method~\cite{\citekohn} for photoionization~\cite{\citekohnphoto} and electron-molecule scattering~\cite{\citekohnelec}.
The photoionization time delays are calculated in two different levels of approximation and then averaged over molecular orientation and outgoing electron direction, consistent with the measurement scheme used in the experiment.
%The first method is referred to as the independent channel calculation and considers the scattering in each continuum channel separately.
The independent channel method considers the scattering in each continuum channel separately.
%The second method uses fully coupled continuum states, which allows electrons originally produced in one ionization channel to interact with the residual ionic core to produce different final state configurations.
The coupled-channel method uses fully coupled continuum states, which allows electrons originally produced in one ionization channel to interact with the residual ionic core to produce different final state configurations.
More details of the complex-Kohn calculation are given below and in the supplemental material~\cite{SM}.

The independent channel calculations~(dashed lines in \Cref{fig::Theory}) for the $X^{2}\Pi_{g}$ and $A^{2}\Pi_{u}$ channels display traditional Coulombic behavior: monotonically increasing photoionization delay with decreasing photoelectron energy~\cite{klunder_probing_2011}.
The $B^{2}\Sigma^{+}_{u}$ channel exhibits an increased photoionization delay time around 25~eV, which is a signature~\cite{baykusheva_theory_2017,hockett_time_2016,huppert_attosecond_2016} of a weak shape resonance that has been observed in the CO$_2$ absorption spectrum. 
The interchannel coupling drastically alters the predicted photoionization time delays~(solid line in \Cref{fig::Theory}). 
The photoionization time delays predicted for the fully-coupled $X^{2}\Pi_{g}$ continuum become extremely long for low energy~($<20$~eV) photoelectrons.
This increase is caused by coupling of the $X^{2}\Pi_{g}$ continuum to Rydberg states converging to the $A^{2}\Pi_{u}$, and $B^{2}\Sigma^{+}_{u}$ state thresholds, i.e. autoionization.
Coupling among the continuum channels also results in a decrease in the photoionization time delay in the vicinity of the shape resonance feature in the $B^{2}\Sigma^{+}_{u}$-state channel.
This decrease is accompanied by an increase in the photoionization time delays in the other channels.

\Cref{fig::XState,fig::ABState} compare our extracted photoionization delays to the theoretical predictions of both models and show that the measured photoionization delays are consistent with the coupled-channel theory.
Moreover, there is strong disagreement with the single-channel predictions in the vicinity of the $B$-state shape resonance.   
These time-domain measurements show how electron correlation dynamics accelerate the escape of the photoelectron from the molecular potential.
Electron interactions cause the EWP in the $B$-state continuum to transition to other available continua while it is trapped in the vicinity of the ionic core.
These transitions produce photoelectrons in the $X^{2}\Pi_{g}$, $A^{2}\Pi_{u}$, and $C^2\Sigma_{g}$ continua with increased photoionization time delays.
The couplings to the continuum channels act as additional pathways for the electron to escape the shape resonance and thus lower the photoionization delay times for the $B^{2}\Sigma^{+}_{u}$ state.

\Cref{fig::XState} shows the measured photoionization time delays for the $X^{2}\Pi_{g}$ channel along with the single-channel and coupled channel calculations.
For photoelectron energies above 20~eV, the measured delays are consistent with both the single-channel and coupled-channel predictions. 
Below 20~eV, the CO$_2$ absorption spectrum displays a series of sharp peaks associated with two Rydberg series converging to the $A^{2}\Pi_{u}$, $B^{2}\Sigma^{+}_{u}$ ionization thresholds~\cite{chan_electronic_1993}.
The observed photoionization time delay confirms the autoionizing nature of the Rydberg states, as shown on the left side of \Cref{fig::XState}.

%%%%%%% End Discussion of Data %%%%%%%%%%%

%%%%%%% Begin Description of Technique %%%%%%%%%%%
\begin{figure}
    \centering
    %\resizebox{\columnwidth}{!}{\includegraphics{spectrograph.eps}}
    \resizebox{\columnwidth}{!}{\includegraphics{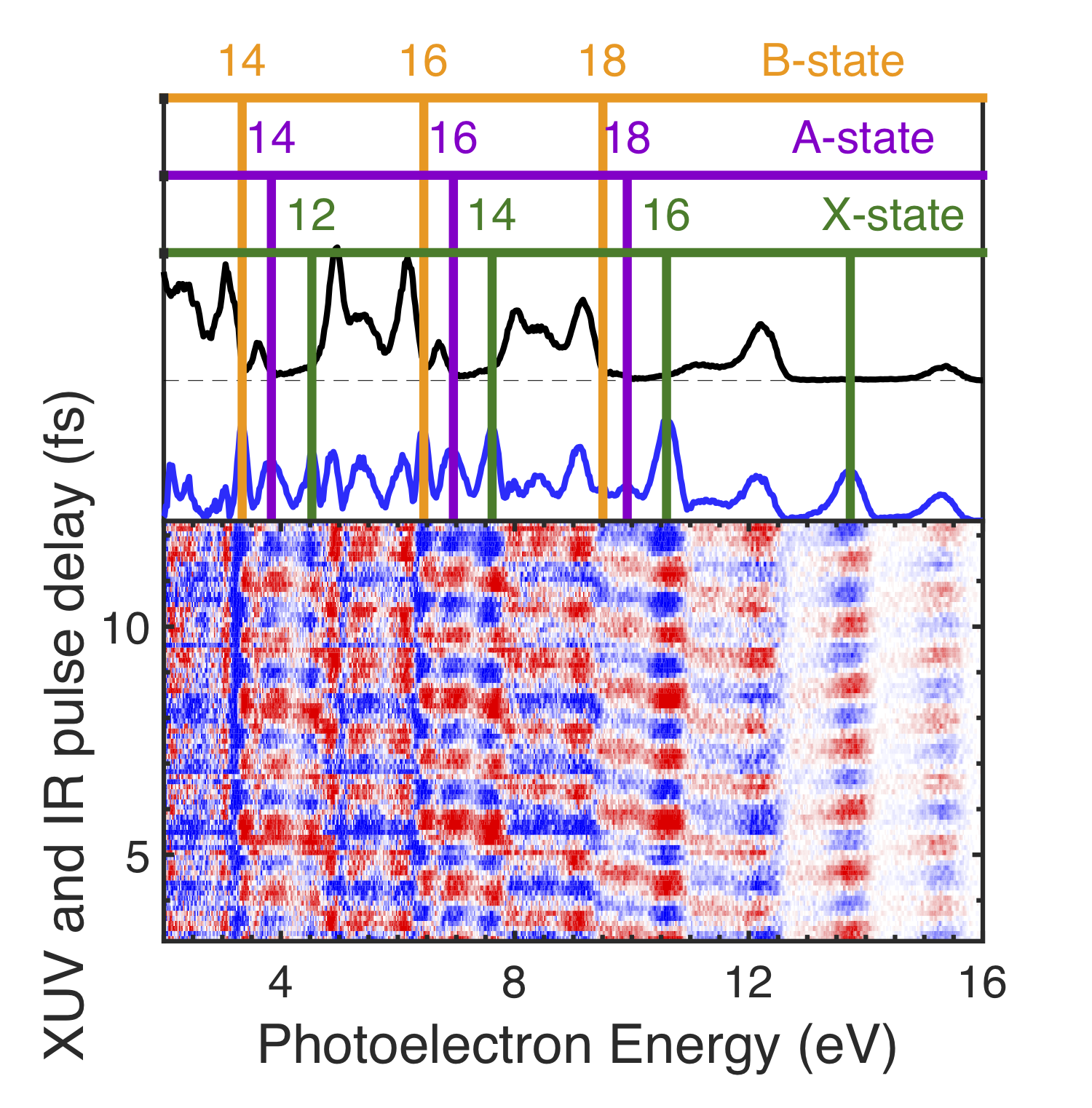}}
    \caption{Two-Color photoelectron spectrogram of CO$_2$.  The upper panel shows the photoelectron spectrum recorded with only the XUV APT present.  The middle panel shows the amplitude of the $2\omega$ oscillation retrieved from the Fourier transform of the lower panel.  The lower panel shows a subset of the measured photoelectron spectrogram and demonstrates clear $2\omega$ oscillation.  All features in this panel have been assigned to either a harmonic or sideband peak of a CO$^+_2$ final state.}
    \label{fig::RawData}
\end{figure}
The measured photoionization time delays displayed in \Cref{fig::dataWithTheory} are obtained using a two-color, multi-path interference technique, which overlaps a weak infrared~(IR) laser pulse, with photon energy $\hbar\omega$, and an XUV frequency comb with spectral peaks separated by $2\hbar\omega$. 
Such a comb appears in the time domain as an attosecond pulse train~(APT), whose pulses provide the temporal resolution needed to measure EWP delays.
To produce this pulse arrangement, a titanium-doped:sapphire laser~(30~fs, 30~mJ, 100~Hz repetition rate) is split into three co-propagating beams: two temporally overlapped, high energy~($\sim14$~mJ) beams used to drive high harmonic generation~(HHG) and produce the XUV APT, and a low energy~($\sim1.5$~mJ) beam used as an interferometric probe. 
All three beams are spatially displaced and focused by a common $f=5$~m focusing optic.
Near the focus, the three beams intersect in a 10~mm long gas cell filled with $\sim7$~torr of argon gas.
A temporal advance~($\sim150$~fs) is introduced in the probe beam path so that it does not disrupt the HHG process when passing through the gas cell.  
The crossed-beam geometry separates the XUV and IR pulses in the far field~(See SM Fig.~S1)~\cite{SM}.  
The residual drive laser light is blocked downstream from the gas cell and the probe beam passes through a $100~\mu$m fused silica window to temporally overlap with the XUV APT. 
%We generate APTs of XUV radiation by frequency upconversion of ultrashort~(30~fs), near-infrared laser pulses~(30~mJ, 100~Hz repetition rate) via high-order harmonic generation~(HHG) in a 10~mm long gas cell, filled with $\sim7$~torr of argon gas.
%Two copropagating, spatially displaced, replica pulses are created upstream of a 5~m focusing optic.
%Crossing the focused beams in the argon gas cell leads to generation of APTs, and this crossed-beam geometry allows for spatial separation of the XUV and IR pulses in the far field~(See SM Fig.~S1)~\cite{SM}. 
%In addition to creating two identical HHG-drive laser pulses, a small amount~($\sim5\%$) of the drive laser is also focused through the gas cell in advance of the drive laser pulse.
%This pulse is later used as the phase-controlled IR pulse for the interferometric measurements.
%At the focusing optic, this beam is also spatially separated from the drive laser pulses, which means that in the far field, this pulse is spatially separated from the XUV pulse.
%To temporally overlap the XUV APT and weak IR pulse, the IR pulse passes through a 100~$\mu$m fused silica window.
The temporal delay of the weak IR field is controlled with a piezo-electric driven delay stage.
Both the XUV APT and weak IR laser pulse are focused with a B$_4$C coated focusing optic~($f=10$~cm) into the interaction region of a 1.2~m magnetic bottle spectrometer~\cite{mucke_hitherto_2010}.
The CO$_2$ target is introduced through a $35~\mu$m gas needle near the interaction region.
The XUV-only photoelectron spectrum of CO$_2$ is shown in the top panel of Fig.~\ref{fig::RawData}, and displays peaks spaced by twice the photon energy of the fundamental drive laser. 
The introduction of the weak IR field allows for two-photon absorption and produces ``sideband'' features between adjacent harmonic spectral features.
Electron spectra are recorded as a function of XUV/IR delay, resulting in modulation of the sideband features as seen in the bottom panel of Fig.~\ref{fig::RawData}.
This modulation occurs at twice the IR laser frequency and can be described by
\begin{equation}
    \text{SB}_{2n}(\tau)\propto\cos\left[2\omega\tau+\phi_{2\omega}\right]
\end{equation}
where $\text{SB}_{2n}$ is the yield of a sideband peak as a function of $\tau$, the relative delay between the IR and XUV pulses, and $\phi_{2\omega}$ is a phase offset in the sideband modulation that varies with sideband order~($2n$).
The phase offset, $\phi_{2\omega}$, is recovered from the Fourier transform of the photoelectron spectrogram.%, which is averaged over the frequency range of the sideband oscillation.
The extracted phase offsets are shown in the supplemental material~\cite{SM}. 

 The sideband modulation is caused by interference between two different ionization pathways formed by single-photon XUV ionization by adjacent harmonics followed by subsequent absorption or emission of an IR photon~\cite{paul_observation_2001,muller_reconstruction_2002,mairesse_frequency-resolved_2005,dahlstrom_introduction_2012}.
The sideband phase offset can be parsed into two contributions:
\begin{equation}
    \phi_{2\omega} = \Delta\phi_{\text{XUV}} + \Delta\phi,
    \label{Eqn::phi_2w}
\end{equation}
where $\Delta\phi_{\text{XUV}}$ describes the spectral phase difference between consecutive harmonics that contribute to the sideband peak, and $\Delta\phi$ describes the phase difference between the two-photon ionization pathways~\cite{muller_reconstruction_2002,dahlstrom_introduction_2012}.
The reconstruction of attosecond bursts by two-photon transitions~(RABBITT) technique was originally developed to characterize APTs by extracting the first term in Eqn.~\ref{Eqn::phi_2w}~\cite{paul_observation_2001,muller_reconstruction_2002}.
Subsequent work has focused on the latter quantity in Eqn.~\ref{Eqn::phi_2w} to approximate a delay for the two-photon ionization process~($\tau_{2}$): 
\begin{equation}
    \frac{\Delta\phi}{2\hbar\omega} \approx \frac{\partial \phi}{\partial E} = -\frac{\tau_{2}}{\hbar}.
    \label{Eqn::2photonDelay}
\end{equation}
In most cases, this two-photon delay can be separated into a measurement induced~(or continuum-continuum) contribution~($\tau_{\text{cc}}$) that simply depends on the energy of the outgoing electron~($\epsilon=E_{2n}-I_p$), and a potential-dependent term~($\tau_{PI}$)~\cite{baykusheva_theory_2017,dahlstrom_introduction_2012, klunder_probing_2011,dahlstrom_theory_2013}:
\begin{equation}
    \tau_2(E_{2n}) \approx \tau_{PI}(E_{2n}) + \tau_{\text{cc}}(\epsilon).
    \label{Eqn::2wDelay}
\end{equation}
When the system is spherically symmetric and the ionization process is dominated by a single angular momentum partial wave, the potential dependent term can be shown to approximate the %dispersion of the scattering phase, $\tau_{PI}=\partial \delta / \partial \epsilon$
single-photon photoionization time delay%
~\cite{dahlstrom_introduction_2012,dahlstrom_theory_2013}.
In this limit, the RABBITT technique has been used to investigate photoionization time delays for different continuum channels in atomic targets~\cite{klunder_probing_2011,isinger_photoionization_2017,jordan_spin-orbit_2017} as well as the relative photoionization time delay between atomic targets~\cite{palatchi_atomic_2014,guenot_measurements_2014}.
The RABBITT technique can also be used to observe resonant processes in atomic photoionization~\cite{kotur_spectral_2016,gruson_attosecond_2016,busto_timefrequency_2018}.

The application of this interferometric technique to molecular systems is more challenging.
Molecular targets often have several accessible cationic states that lead to substantial overlap of features in the photoelectron spectra~(spectral congestion)~\cite{jordan_extracting_2018}.
Moreover, the partial-wave expansion of outgoing photoelectron wavepackets can contain a large number of coherent contributions, a challenge not typically encountered in atomic targets~\cite{fano_propensity_1985}.
Nevertheless, Huppert~\textit{et al.} recently observed the effect of a molecular shape resonance on the measured photoionization time delays in N$_{2}$O$^{+}$~\cite{huppert_attosecond_2016}.
Vos~\textit{et al.} were able to study the stereo Wigner time delay in carbon monoxide averaged over a number of dissociative states of the CO$^+$ cation~\cite{vos_orientation-dependent_2018}. 
Due to the excellent kinetic energy resolution afforded by the magnetic bottle spectrometer, we are able to resolve the sideband oscillations for three final state channels in CO$_2$~(middle panel of \Cref{fig::RawData}) and compare these results with theory predictions.

The relevant term in Eqn.~\ref{Eqn::2wDelay} for theory comparison is the potential-dependent term, $\tau_{\text{PI}}$.
To extract this contribution from the measured phase offsets, we consider the phase differences between a signal~$(s)$ and reference~$(r)$ channel:
\begin{eqnarray}
    \label{Eqn::RelRABBITT}
    \Delta\tau_2(E_{2n}) &=& \frac{-1}{2\omega}\left[\phi^{(s)}_{2\omega} - \phi^{(r)}_{2\omega}\right] \\
    & = & \Delta\tau_{\text{PI}}(E_{2n}) + \Delta\tau_{cc}(\epsilon,\epsilon^{\prime}), \nonumber
\end{eqnarray}
where $\epsilon=E_{2n}-I_p^{(s)}$ and $\epsilon^{\prime}=E_{2n}-I_p^{(r)}$ are the kinetic energies of the photoelectrons for the $2n^{\text{th}}$ sideband peak in the signal and reference systems, $\Delta\tau_{cc}(\epsilon,\epsilon^{\prime})$ is the difference in the continuum-continuum contribution due to the mismatch in photoelectron energies and $\Delta\tau_{\text{PI}}(E_{2n})$ is the differential photoionization delay we will compare with calculations.
The spectral phase variation of the XUV pulse train, $\Delta\phi_{\text{XUV}}$, has canceled out. 
When the relative difference between $\epsilon$ and $\epsilon^{\prime}$ is small, $\Delta\tau_{cc}(\epsilon,\epsilon^{\prime})$ can be calculated very accurately~\cite{dahlstrom_theory_2013}, and subtracted from Eqn.~\ref{Eqn::RelRABBITT}.
For the measurements presented in \Cref{fig::ABState} we reference the $B$-state photoionization delay to that of the $A$-state because the two channels have similar ionization potentials~($I_{p,A} = 17.59$~eV and $I_{p,B} = 18.07$~eV).
The $X$-state channel ($I_{p,X} = 13.77$~eV, \Cref{fig::XState}) is compared with a reference measurement made in krypton gas ($I_{p,\text{Kr}} = 14.00$~eV).
The krypton target is well studied~\cite{jordan_spin-orbit_2017} and its photoionization delay has been calculated~\cite{kheifets_time_2013}, so we remove this contribution in the differential photoionization delay.

We calculate the photoionization time delay starting from the time evolution of the EWP. 
In the time-domain, the EWP,~$\psi(t)$, is expressed as a coherent superposition of outgoing continuum electron eigenstates~($|\psi_{\vec{k},n}^{(-)}\rangle$) with momentum $\vec{k}$:
\begin{equation}
    \left|\psi(t)\right\rangle = \sum_n \int d\vec{k}~c_n(\vec{k})~e^{-\frac{it}{\hbar}\left(\frac{k^2}{2}-I_{p,n}\right)}~ \left|\psi_{\vec{k},n}^{(-)}\right\rangle,
    \label{Eqn::EWP}
\end{equation}  
where $n$ describes the final state channel with ionization potential $I_{p,n}$, and $|c_n(\vec{k})|^2$ is the photoionization probability.
Additionally, we define the scattering phase $\delta_n(\vec{k}) = \text{Arg}[c_n(\vec{k})]$. 
The form of the outgoing electron wavefunction is described in the complex Kohn formalism and is discussed fully in the supplemental material~\cite{SM}. 
%The outgoing electron wavefunction is a multi-electron ($N$) wavefunction:
%The outgoing electron wavefunction is commonly expressed as an anti-symmetrized product of $N−1$ bound electronic wavefunction with a single free electron wavefunction, 
%wavefunction:
%\begin{equation}
%    \psi_{\vec{k},n}^{(-)} &=& \sum_m\left[ \hat{A}\left\{\Phi_m(r_1,...,r_{N-1})\phi^{(-)}_{nm}(r_N)\right\}\right] 
%\end{equation}
%where $\Phi_n$ is the anti-symmetric, multi-electron wavefunction of the ionic core, $\hat{A}$ asymmetrizes the product, and $\phi^{(-)}_{nm}$ is a single-electron wavefunction satisfying the scattering boundary conditions~\cite{newton_formalism_1982}.
%This function can be expressed asymptotically as,
%\begin{equation}
%    \phi_{nm}(r) \xrightarrow{r\rightarrow\infty} \left(\delta_{nm}h^{(+)}_{\vec{k},n}(r_N)+S_{nm}^*(\hat{k})h^{(-)}_{\vec{k},n}(r_N)\right),
%\end{equation}
%where $h^{(+)}_{\vec{k},n}$ is the regular-incoming Coulomb function, $h^{(-)}_{\vec{k},n}$ is the irregular-outgoing Coulomb function, and $S_{nm}$ is the $S$-matrix of the electron scattering problem.   
%
Initially~(at $t = 0$) the EWP is localized near the ionic core, but it is not stationary.
The phase of each eigenstate component,
\begin{equation}
    \Phi_n(\vec{r},\vec{k}, t) = \text{Arg}\left[\psi_{\vec{k},n}^{(-)}(\vec{r})\right]+\delta_n(\vec{k})-\frac{t}{\hbar}\left(\frac{k^2}{2}-I_{p,n}\right), 
    \label{Eqn::Phase}
\end{equation}
evolves with time, and the wavepacket moves and disperses.
%It  moves and disperses as the phases of its eigenstate components evolve with time.%with the time propagator $e^{i(-\frac{k^2}{2}+I_{p,n})\frac{t}{\hbar}}$. 
%The spatial location of maximum constructive interference also evolves, signifying the movement of the EWP. 
The spatial location of maximum constructive interference is given by the stationary points of Eqn.~\ref{Eqn::EWP}:
\begin{equation}
    \left.\frac{\partial\Phi_n(\vec{r},\vec{k}, t)}{\partial\vec{k}}\right.=0.
    \label{Eqn::Stationary}
\end{equation}
%Asymptotically~($r\rightarrow\infty$), $\psi_{\vec{k},n}^{(-)}$ can be expanded in a basis of % multi-electron cationic wavefunctions and single-electron Coulombic wavefunctions.
%product wavefunctions~($N-1$ electron bound-state wavefunctions with a single-electron Coulombic wavefunction).
%, and the stationary point of Eqn.~\ref{Eqn::EWP} will have the form,
%\begin{equation}
%\frac{\partial\Phi_n}{d\vec{k}}=\vec{r} - %\frac{\text{ln}2\vec{k}\cdot\vec{r}}{k^2}+\frac{1}{k^2}+\frac{\partial\delta_n}{\partial\vec{k}}-%\frac{\vec{k}t}{\hbar}=0.
%    \label{Eqn::ClassicalMotion}
%\end{equation}
%also evolves, signifying the movement of the EWP.

An observable delay time, $\tau_1$, measures the time at which an EWP of energy $E=\nicefrac{k^{2}}{2}$ arrives at a detector some fixed position $\vec{r}$~($|\vec{r}|\gg1$) from the origin, compared to the case of no scattering~\cite{dahlstrom_introduction_2012}.
Combining Eqn.~\ref{Eqn::Phase} and~\ref{Eqn::Stationary}, this direction-specific value can be shown to be
\begin{equation}
    \tau_1 =-\frac{\hbar}{k}\frac{\partial \delta_{n}(\vec{k})}{\partial k} =-\hbar\frac{\partial \delta_{n}(\vec{k})}{\partial E}.
    \label{Eqn::EWS}
\end{equation}
%Eqn.~\ref{Eqn::EWS} then implies that the arrival time of the EWP at the detection position will be affected by changes in the scattering phase of the ionic potential.  
Eqn.~\ref{Eqn::EWS} shows the observed time delay is a direct consequence of the scattering phase, which depends on the charge dynamics that occur during photoionization.
$\tau_1$ is commonly referred to as the single-photon photoionization time delay~\cite{dahlstrom_introduction_2012,baykusheva_theory_2017} or simply as the Eisenbud-Wigner-Smith delay, $\tau_{\text{EWS}}$~\cite{Wigner1955LowerShift}.
The measurements and calculations presented in figure \ref{fig::dataWithTheory} are related to $\tau_1$, but the presence of the dressing laser field, along with the experimental measurement geometry can complicate the relationship~\cite{baykusheva_theory_2017,hockett_time_2016}.  
%based on estimating the value $\frac{\partial \delta_{n}(\vec{k})}{\partial E}$ and integrating over all possible directions $\vec{k}$ and polarization directions $\vec{\mu}$.
The delays retrieved from the RABBITT technique, $\tau_{\text{PI}}$, and $\tau_1$ become identical in the special case of a spherically symmetric potential, when a single angular momentum channel dominates the photoionization process.   
%$\delta_{n}(\vec{k}) = \delta_{n}(k)$ this value defines the Eisenbud-Wigner-Smith delay $\tau_{EWS}$.
%In one of the essential theorems connecting time-dependent and time-independent descriptions of scattering – relating the properties of the time-independent Green’s function with the time-dependent Moller operator~\cite{newton_formalism_1982} - the Wigner time delay, given by $\partial_{\epsilon}\delta(\epsilon=\nicefrac{k^2}{2})$, is shown to be proportional to the delay or latency time experienced during the scattering process, the degree to which the maximum of the departing wave packet is delayed relative to the case of no scattering.
%The evolution of the EWP is governed by the dispersion of the scattering phase,$\tau_{W}=\partial_{\epsilon}\delta(\epsilon=\nicefrac{k^2}{2})$. 
%The Wigner time delay~\cite{Wigner1955LowerShift}, given by $\tau_{W}=\partial_{\epsilon}\delta(\epsilon=\nicefrac{k^2}{2})$, is shown to be proportional to the delay or latency time experienced during the scattering process. 
%$\tau_{W}$ describes the degree to which the maximum of the departing wave packet is delayed relative to the case of no scattering~\cite{Wigner1955LowerShift, dahlstrom_introduction_2012}.

%For weak XUV pulses, the ionization amplitude can be calculated from first order perturbation theory per Eqn.~\ref{Eqn::DME}.
%Eqn.~\ref{Eqn::DME} describes the first-order perturbation theory result for the molecular ionization amplitude from weak XUV pulses. 
The ionization probability amplitude in Eqn.~\ref{Eqn::EWP} can be calculated for weak XUV pulses from first order perturbation theory:
\begin{eqnarray}
    c_n(\vec{k}) &=& \Tilde{E}_{\text{XUV}}\left(\omega = \frac{k^2+2I_{p,n}}{2\hbar}\right)\cdot d_n(\vec{k}) \\
    d_n(\vec{k}) &=& \left.\left.\left\langle\psi_{\vec{k},n}^{(-)}\right|\hat{\mu}\cdot\vec{r}\right|\Psi_0\right\rangle,
    \label{Eqn::DME}
\end{eqnarray}
where $\Tilde{E}_{\text{XUV}}(\omega)$ is the Fourier transform of the incident XUV pulse, $d_n(\vec{k})$ is the dipole matrix element between the outgoing and ground state~($\Psi_0$) wavefunctions in the length gauge, and $\hat{\mu}$ is the polarization direction in the molecular frame.
The channel-resolved dipole matrix element, Eqn.~\ref{Eqn::DME}, is computed using the complex Kohn method for photoionization~\cite{\citekohnimp,\citekohnphoto}.
The calculations use explicit representations of the initial neutral state and of the final cationic states obtained with one single 11-orbital basis.
This basis for the neutral and cationic states was obtained using a state-averaged multiconfiguration self-consistent-field~(MCSCF) calculation performed with the \textit{COLUMBUS} quantum chemistry program~\cite{Lischka_New_2009,Shepard_Progress_1988,Lischka_High_2001,Lischka_Columbus_2011,Coumbus7}.
%For this state-averaged calculation, the energy of the ground neutral and the six cation states was minimized, with relative weights 5:1:1:1:1:1:1. 
The orbitals obtained are a compromise between those optimized for the neutral and cationic states.
The primitive basis for this MCSCF calculation was Dunning’s \textit{aug-cc-pVDZ} basis set~\cite{Dunning_Gaussian_1989}, with additional basis functions on the oxygen atom as described in the supplemental material~\cite{SM}.
%Partial-wave scattering eigenfunctions $\psi^{(-)}_{lm,n}$ are constructed according to the complex Kohn variational principle using partial waves 
The outgoing wavefunction, $\psi^{(-)}_{n}$, is expanded in a partial-wave representation~($l,m$), up to $l=3$, and the dipole matrix element is calculated between these functions and the initial ground state within the approximation of separable exchange~\cite{Rescigno_Separable_1981,Rescigno_Disappearance_1988,McCurdy_Beyond_1992}.  
%More details of these calculations are given in the supplemental material~\cite{SM}.

All meaningful partial waves are then coupled by the IR dressing field as described in Ref.~\cite{baykusheva_theory_2017} to determine the molecular frame~(MF) two-photon photoionization time delays.
% $\tau_{mol}(\epsilon,\hat{k},\hat{R})$, in which $\hat{k}$ describes the photoelectron emission direction in the molecular frame, and $\hat{R}$ describes the orientation of the molecular axis to the XUV polarization direction.
These MF photoionization time delays are then averaged over both the polarization direction and outgoing electron direction to approximate the experimental conditions.
This averaged quantity is what we refer to as the laboratory-frame~(LF) photoionization delay as defined in Eqn.~\ref{Eqn::2wDelay} and plotted in \Cref{fig::dataWithTheory}.
% $\tau_{\text{PI}}$.
%However, t
%The interferometric method for measuring photoemission time delays makes only a finite difference approximation to the derivative in Eqn.~\ref{Eqn::2photonDelay}, which averages the calculated delay value over the XUV frequency comb.
%This averaging explains why the sharp features expected for the autoionizing resonances in the $X$-state continuum are washed out in the calculated delay.
The single photon LF photoionization delays and two-photon MF photoionization delays for all accessible final state channels are shown in the supplemental material~\cite{SM}.

%All populated partial waves are coupled in the molecular frame (MF) as described in the supplemental material~\cite{SM} and in Ref.~\cite{baykusheva_theory_2017} to account for the effects of the IR dressing field.
%The MF photoionization time delays are calculated for all pairs of laser orientation and electron emission angles.
%These time delay values are calculated with the finite difference approximation described in eqn. \ref{Eqn::2photonDelay}.
%The angle dependent MF photoionization delay times are weight-averaged over the complete angle space to approximate the experimental conditions.
%The theory curves presented in figure \ref{fig::dataWithTheory} are these weight-averaged values.
%The MF two-photon photoionization delays for all four energetically accessible cation states are shown in the supplemental material~\cite{SM} as is a comparison of single photon and two-photon photoionization time delays.

Observing phenomena such as autoionization, which typically have lifetimes longer than the inter-pulse spacing of our XUV pulse train~(1.35~fs), requires additional considerations in the analysis.
Since each attosecond burst in the train has a coherent relationship to the neighboring pulses, when time-dependent phenomena extend beyond the inter-pulse spacing, the processes induced by a single burst in the train interfere with the signal generated by the adjacent pulses.
This interference leads to a burst-dependent photoemission time delay, i.e. the peak emission time is the point where all previous photoemission processes interfere constructively.   
Such a modulation~(burst-to-burst) of the photoemission time delay leads to a variation of the phase across a sideband.
This revelation is the motivation behind the development of the Rainbow RABBITT technique~\cite{gruson_attosecond_2016,busto_timefrequency_2018}.
The analysis of the slope of $\phi_{2\omega}$ across each sideband peak is shown in \Cref{fig::Rainbow}, where each channel is referenced to the krypton measurement for the $^2P_{3/2}$ continuum. 
%The krypton reference has non-zero group delay resulting from variation in the XUV pulse train. 
The envelope of the IR-drive laser leads to small time shifts in the arrival time of each burst in the APT~(referred to as harmonic chirp or femto-chirp~\cite{varju__frequency_2005}) which results in linear phase variation across a sideband peak.
This contribution is removed by the krypton referencing because all measurements are made with the same APT.
%Additionally the presence of unresolved vibrational structure in the CO$_2$ measurements could slightly alter the phase variation across a sideband peak and lead to non-zero values in \Cref{fig::Rainbow}.
From the residual slope value obtained for the 12$^{\text{th}}$ harmonic sideband, there is a clear difference compared to any other sideband peak.
Moreover, this difference is consistent with a delay of more than one complete $2\omega$-cycle. 
Instead of relying directly on the value extracted from \Cref{fig::Rainbow}, we use the slope variation to determine the number of cycles to add to the mean phase extracted from averaging the phase over the sideband peak. 

%%%%%%% End Section on Technique %%%%%%

%%%%%%% Conclusions %%%%%%%
These measurements and calculations demonstrate the effect of electron correlation upon time delays in photoionization of molecules.
Photoionization time delays are a direct and easy-to-understand manifestation of the binding interaction that an outgoing electron experiences.
We have observed two effects of electron correlation in time-domain measurements of CO$_2$ photoionization, via (1)~a shape resonance and (2)~autoionizing states, demonstrating that the inclusion of electronic correlation is important when considering resonant features in molecular photoionization with XUV light.
While this seems clear for autoionizing states, which are inherently multi-channel phenomena, this result is somewhat surprising for shape resonance features which are typically considered to be single channel phenomena.
For a shape resonance, we have shown that electron-electron interactions provide dynamic modifications to the effective potential that can be directly observed in all final state channels.
These results highlight the need for including electron correlation when describing time-domain measurements of photoionization of molecular targets, where autoionizing states and shape resonances are omnipresent.
%djh commented 05-07-19  Therefore, the appropriate inclusion of electron correlation is necessary for interpreting photoionization time delay measurements in the vicinity of any resonant feature. 

Molecular control (e.g. molecular alignment) techniques can be applied to isolate specific cationic states of CO$^{+}_{2}$ in future experiments.
Continuum-resolved molecular frame measurements will further elucidate the dependence of electron correlation on molecular orientation and electron emission angle.
These experiments could be further improved by varying the XUV frequency comb spacing, thus mapping out more energy points in the differential scattering phase. 

%%%%%%%% \section*{\label{sec:acknowledgements}Acknowledgements}

This work was supported by the AMOS program within the Chemical Sciences, Geosciences, and Biosciences Division of the Office of Basic Energy Sciences, Office of Science, U.S. Department of Energy.
Raw and analyzed data, both experimental and theoretical, along with a subset of the analysis code is available at:\url{https://figshare.com/account/home#/projects/63164}

%%%%%%%%% \section*{\label{sec:compInt}Competing Interests}

%\bibliography{references.bib}

%\bibliography{referencesJPC.bib,co2_2019_djh_kohn.bib,co2_2019_djh_other.bib}

%outdated commented out version is at%\bibliography{referencesJPC.bib,co2_2019_djh_kohn.bib}

%apsrev4-2.bst 2018-12-27 (MD) hand-edited version of apsrev4-1.bst

\end{document}

\title{Supplementary Material for ``Electron Correlation Effects in Attosecond Photoionization of CO$_2$''}

\author{Andrei Kamalov$^{1,2}$}
\author{Anna L. Wang$^{1,3}$}
\author{Philip H. Bucksbaum$^{1,2,3}$}
\author{Daniel J. Haxton$^{4}$}
\author{James P. Cryan$^{1,5}$}
\affiliation{%
$^{1}$Stanford PULSE Institute, SLAC National Accelerator Laboratory, Menlo Park, CA, USA
}%
\affiliation{
$^{2}$Department of Physics, Stanford University, Stanford, CA, USA
}
\affiliation{
$^{3}$Department of Applied Physics, Stanford University, Stanford, CA, USA
}
\affiliation{
$^{4}$KLA Corporation, Milpitas, CA, USA
}
\affiliation{
$^{5}$Linac Coherent Light Source, SLAC National Accelerator Laboratory, Menlo Park, CA, USA
}

\date{\today}
\maketitle
\tableofcontents

\renewcommand{\thefigure}{S\arabic{figure}}
\setcounter{figure}{0}

\section{Experimental Setup}
\subsection{Optical Laser}
\begin{figure}
    \resizebox{\columnwidth}{!}{
        \includegraphics{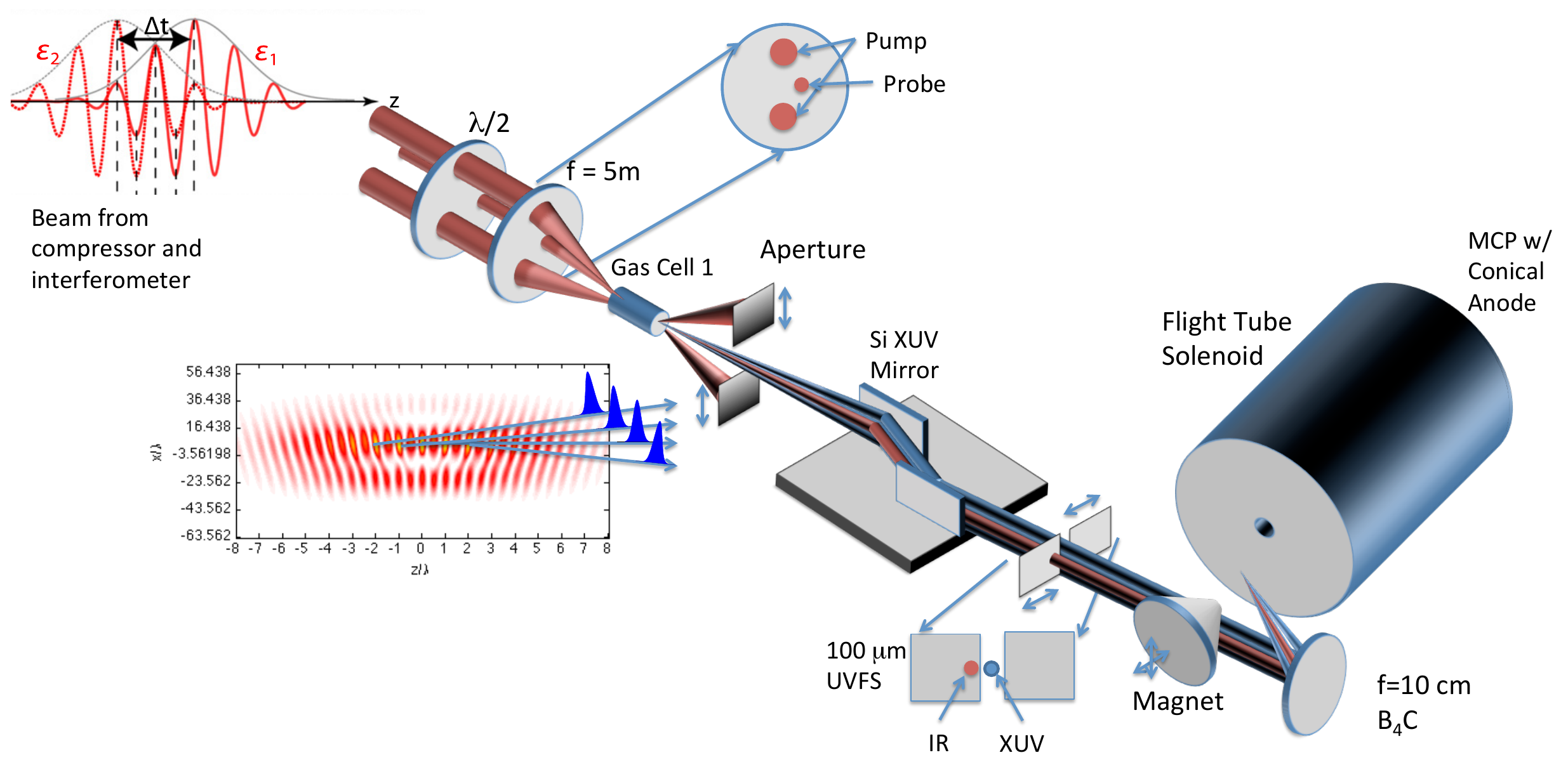}}
    \caption{Experimental Layout showing crossed beam geometery}
    \label{fig::ExpLayout}
\end{figure}

The general layout of the optical system is described in the main text.
Here we supplement the previous description with some additional details. 
The three beams used in the experiment are created by a set of nested interferometers, that produce three spatially separated beams, as shown in Fig.~\ref{fig::ExpLayout}.
We monitor the stability of the pump beam interferometer by measuring the fringe stability in the focused beams. 
The pulse compressor and nested interferometers reside within a low-vacuum chamber (180~Torr) that is back-filled with helium.
The helium atmosphere maintains thermal equilibrium for optical components while minimizing chromatic dispersion and nonlinear propagation effects.
All three beams are focused with a common $f=5$~m focusing optic.
After which, all three beams pass through a thin glass window which separates the low-vacuum chamber from the remainder of the beamline, which is maintained at high vacuum.
As shown in Fig.~\ref{fig::ExpLayout} all three beams intersect near the focus, inside of a $\sim1$~cm gas cell filled with $\sim7$~torr of argon gas. 
The sides of the gas cell consist of replaceable thin copper shims which are laser-drilled through to produce apertures the size of the IR beams.

The XUV pulse train, produced by the overlapping pump beams, propagates between the two pump beams and is spatially filtered from the IR pump beams.
A thin glass plate is introduced to temporally retard the probe field by $\sim$150~fs while simultaneously retarding background IR by the same value~(see Fig.~\ref{fig::ExpLayout}).
This temporally overlaps the XUV APT with the probing laser field. 

As shown in Fig.~\ref{fig::ExpLayout} the XUV and IR-probe beams are back-focused with a B$_{4}$C coated 0$^{\circ}$ focusing optic~($f=10$~cm).
The B$_{4}$C coating reflects up to the 19$^{\text{th}}$ harmonic~(29~eV) of the HHG drive laser.
The focus spot is vertically offset such that the focus is spatially separated from the incoming XUV beam path.
A 35~$\mu$m effusive gas needle is positioned just above the focus spot.
A 1.22~m magnetic bottle type time of flight~(ToF) detector is used to collect photoelectrons produced by the focused XUV beam. 
The gas needle and permanant magnet were positioned outside of the true focus of the IR-field to avoid the Gouy phase shift.

\subsection{Photoelectron Spectrometer}
The magnetic bottle solenoid was set to have a uniform magnetic field of strength $\sim$300~$\mu$T and used a permanent magnet with peak field $\sim$ 1~T.
A soft iron cone was attached to the magnet, and the cone tip was positioned near the interaction region.
The base pressure for the chamber is $4\times10^{-9}$~Torr, and the experiments were performed with a chamber pressure close to $2\times10^{-7}$~Torr. 
A multi-channel plate~(MCP) detector with a conical anode is positioned at the end of the solenoid.
The anode signal is amplified and digitized by a computer mounted digitizer card with 1~ns precision.
The digitized trace is discriminated, and each recorded time-of-flight~(ToF) is saved to disk. 
The resolution of the photoelectron spectrum was used to optimize the positions of the gas needle, the permanent magnet, the magnetic bottle's solenoid orientation, and solenoid current.

\section{Data Analysis}
XUV/IR delay scans were performed for the target, CO$_2$, and reference, Kr, in alternate fashion.
We scan the XUV/IR delay randomly over 223 unique delay values with 120~as spacing.
An additional delay point, where the XUV and IR fields do not temporally overlap, is recorded as an additional background.  
After acquiring the photoelectron spectrum for 1400~laser shots at each delay point~($\sim40$~min.), the data is saved to disk, and the gas target is changed. 
The data presented in Fig.~1 is the result of 51 independent measurement scans.
The reference gas data enables photoelectron energy calibration, delay-stage drift compensation, and provides a reference signal for the $X^{2}\Pi_{g}$ state data.

\subsection{Extracting the $2\omega$ signal}
\begin{figure}
    \centering
    \resizebox{\columnwidth}{!}{\includegraphics{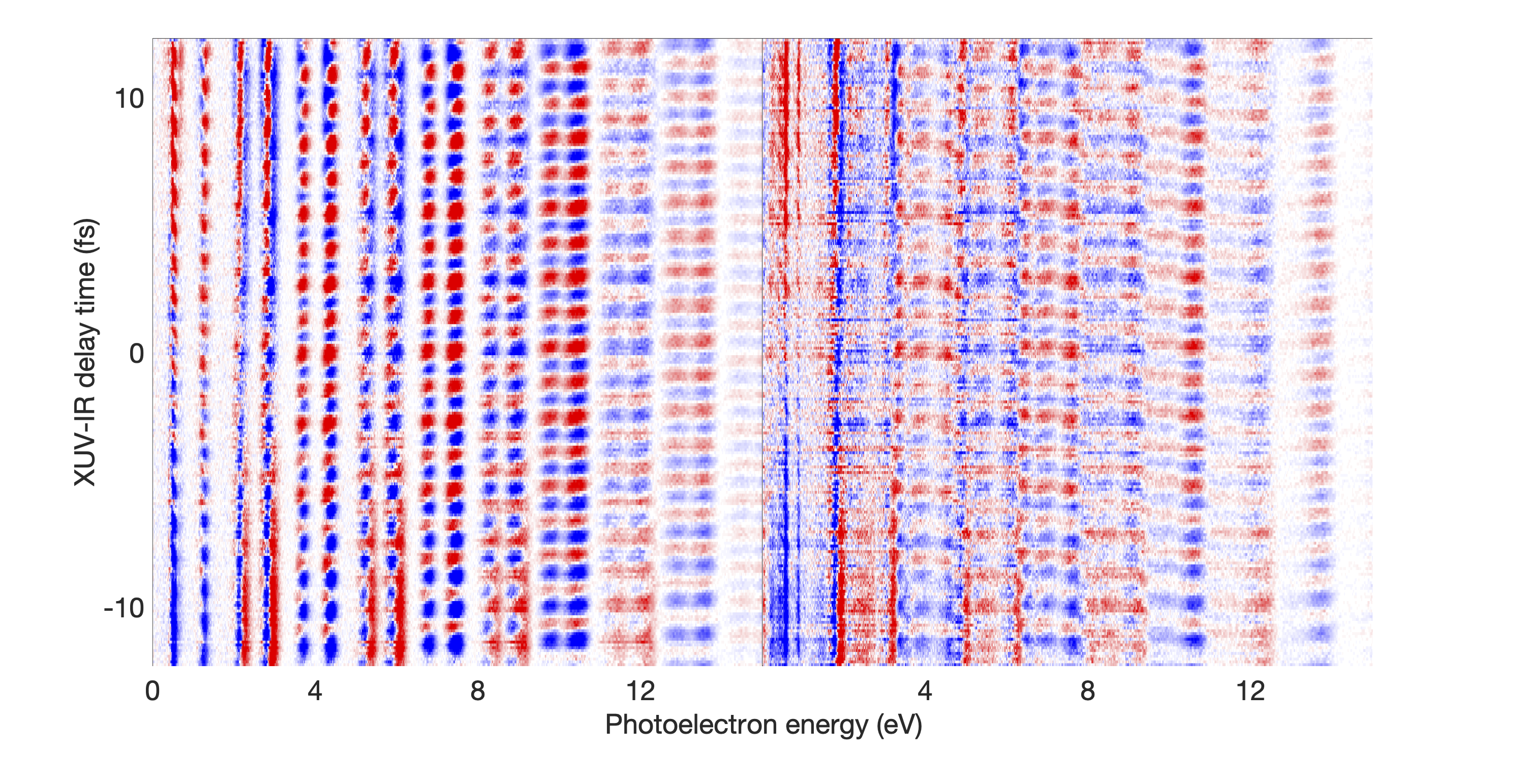}}
    \caption{Spectrographs for Kr (left) and CO$_2$ (right).  Photoelectron spectra are collected at many different time delay values between the XUV and IR fields.  As the fields are varied, there is a clear beating in the amplitude of the photoelectronic signal.  The spectrographs shown here are mean-subtracted for visual clarity.}
    \label{fig:bothSpectrographs}
\end{figure}
Fig. \ref{fig:bothSpectrographs} shows the delay-dependent photoelectron spectra collected for both target gases.
The figures show that a clear beating signal is observed in the photoelectron spectra as the time delay is varied.
The sideband phase offset from Eqn.~1 of the main text is isolated by Fourier transforming the photoelectron spectrogram, and extracting the phase of the $2\omega$ component.
This phase is then averaged over a 0.1~eV~(0.3~eV) wide window centered on each sideband peak of CO$_2$~(Kr), weighted by the Fourier Amplitude~(spectral integration).
\begin{equation}
    \text{Arg}\left[\int A_{2\omega}(E)e^{i\phi_{2\omega}(E)} dE\right]
\end{equation}
Spectral fitting was performed to test whether the $A ^{2}\Pi_{u}$ and $B ^{2}\Sigma^{+}_{u}$ signals were affected by the $C ^{2}\Sigma^{+}_{g}$ state.
The spectral fitting results matched the values returned from spectral integration within confidence intervals.

\subsection{Delay-stage drift compensation}
\begin{figure}
\centering
\resizebox{0.4\columnwidth}{!}{\includegraphics{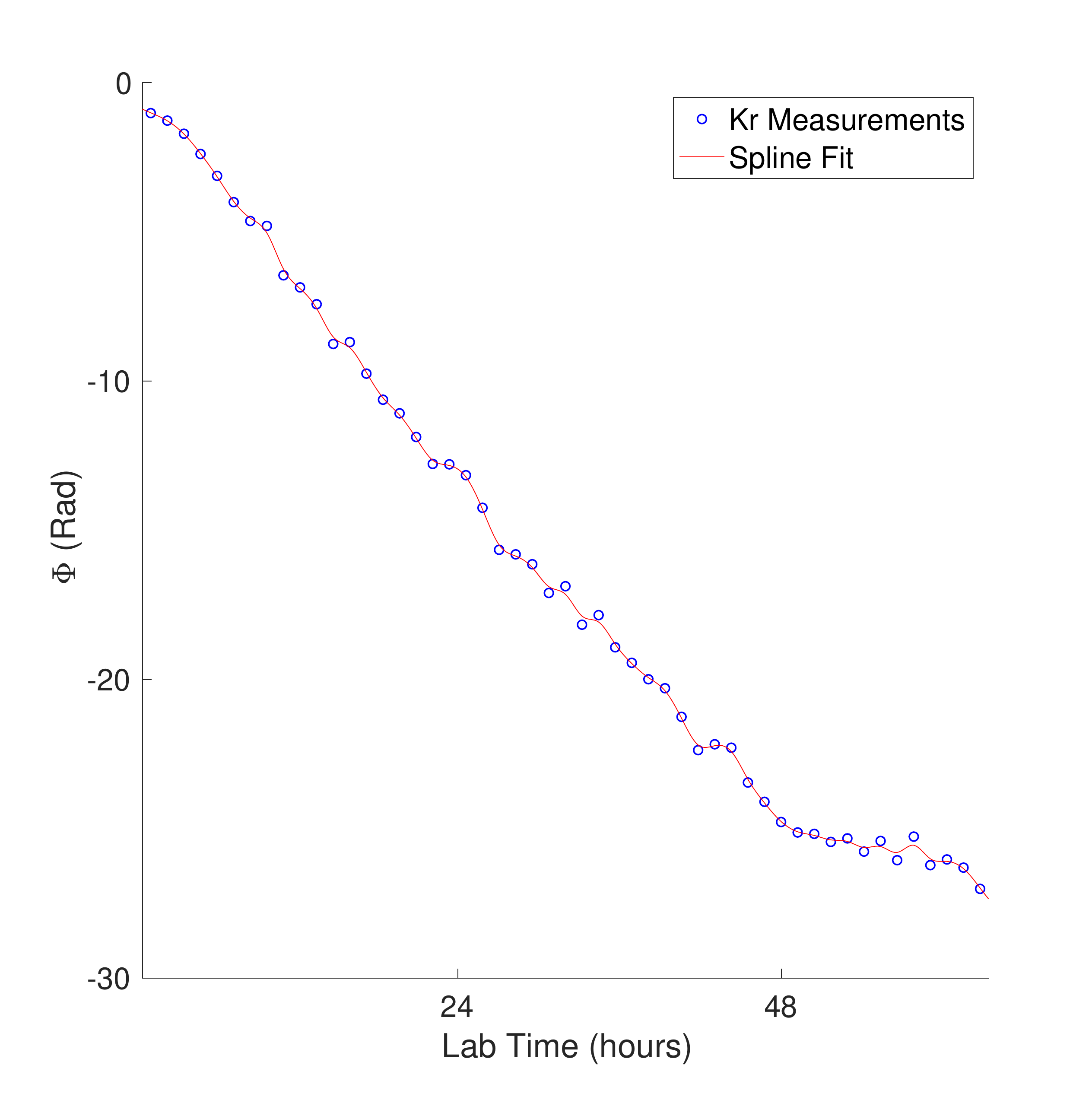}}
\caption{Variation of $\Phi$, the delay-stage drift metric, throughout the duration of the run.  Measurements (blue) are made during Kr scans.  A spline fit (red) is used to estimate delay-stage drift values during CO$_2$ scans performed between adjacent Kr scans.}
\label{fig:delayDrift}
\end{figure}
Comparing the phase offset across the 51 independent measurement scans, requires precise knowledge of any drifts in the XUV/IR laser delay.  
We use the interleaved measurements of the krypton reference to track the long-term drift of the delay stage.
For each Kr scan, the extracted sideband offset from spectral integration is combined into a weighted average,
\begin{equation}
\Phi = \sum_{n = SB} \text{Arg}[ e^{i\cdot\phi_{2\omega}(E_{n})}],
\label{eqn:stageDriftMetric}
\end{equation}
which is used to track the delay-stage drift across the length of the run.
Figure \ref{fig:delayDrift} shows the evolution of $\Phi$ throughout the duration of the experimental run.
A smoothing spline is fit to the experimental values of $\Phi$ and is used to interpolate the value of $\Phi$ at each CO$_2$ scan.  
We subtract the interpolated value of $\Phi$ from each CO$_2$ scan to account for delay-stage drift.  
This procedure accounts for most, but not all of the delay-stage artifact.

\begin{figure}
\centering
    \begin{subfigure}{0.325\columnwidth}
    \resizebox{\columnwidth}{!}{\includegraphics{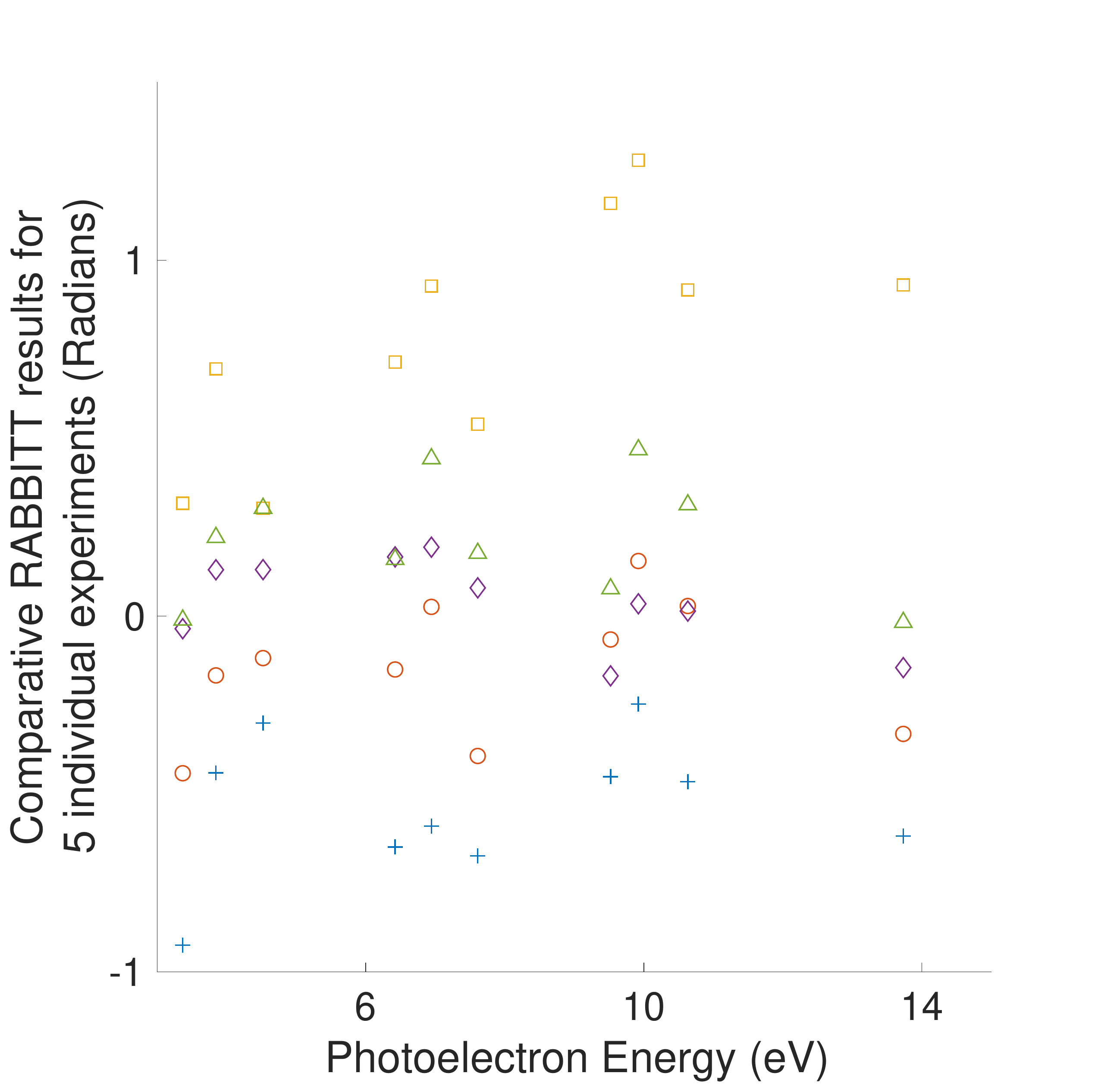}}
    \phantomsubcaption \label{fig:fiveComparisons}
    \end{subfigure}
    \begin{subfigure}{0.325\columnwidth}
    \resizebox{\columnwidth}{!}{\includegraphics{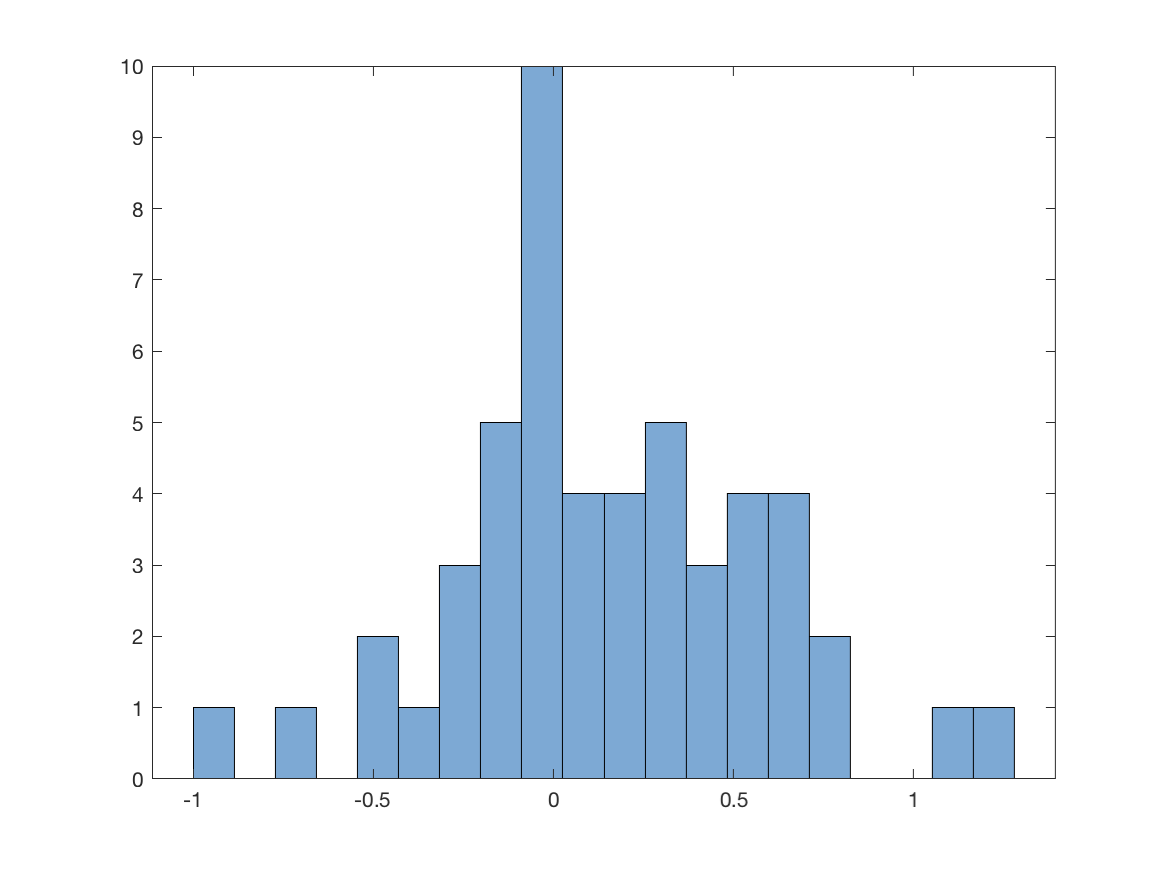}}
    \phantomsubcaption\label{fig:phaseOffsetDistribution}
    \end{subfigure}
    \begin{subfigure}{0.325\columnwidth}
    \resizebox{\columnwidth}{!}{\includegraphics{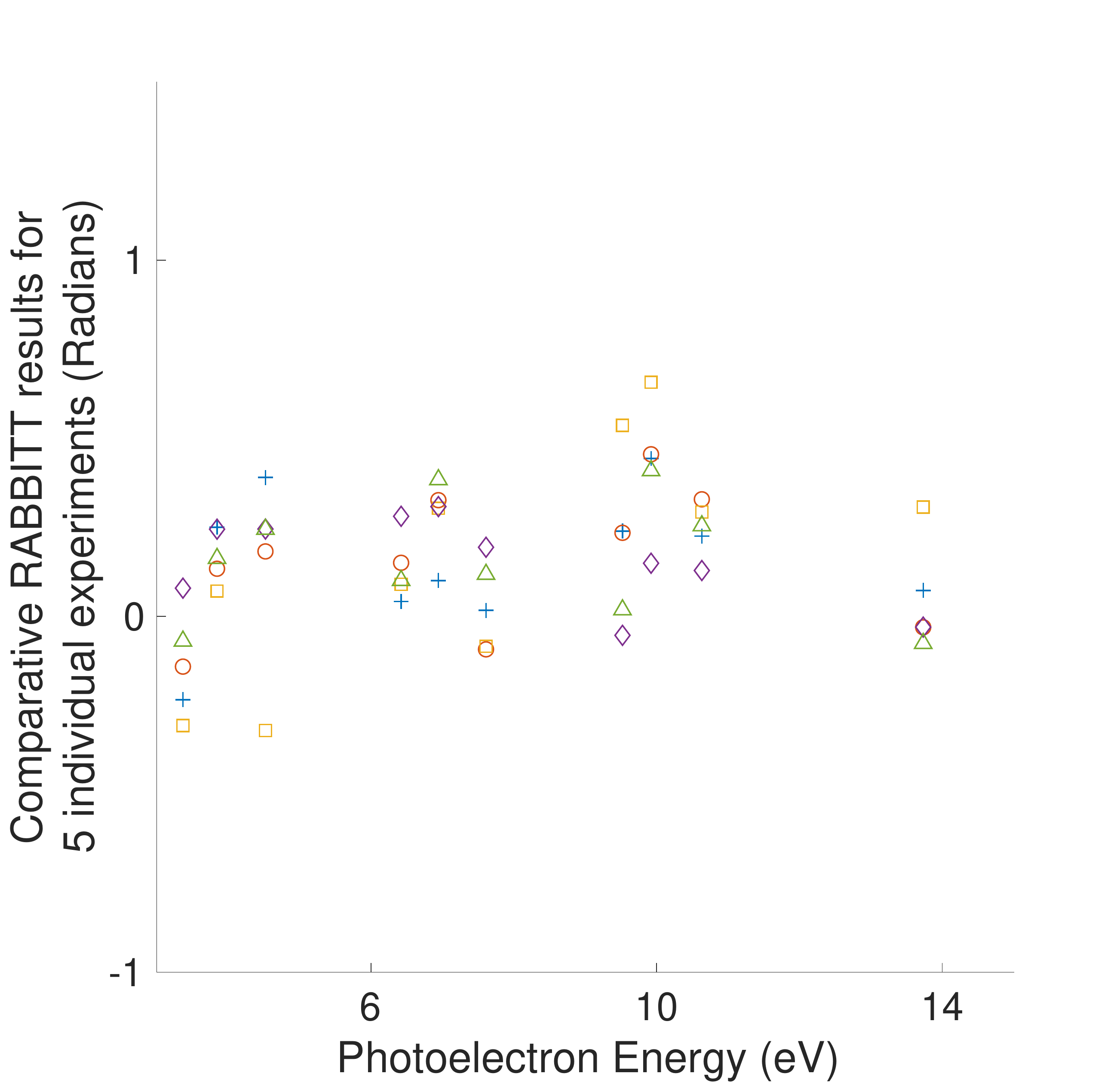}}
    \phantomsubcaption\label{fig:fiveComparisonsPostOffset}
    \end{subfigure}
    \caption{a) (left): the comparative APTS results for five chronologically simultaneous experiments.  These results are shown after the delay-stage drift values are compensated by interpolating drift between Kr reference measurements.  It is clear that there is still a systematic error in timing overlap, given that all scans showcase similar trend lines but with different offsets. b) (middle): a histogram distribution of the averaged phase offsets measured. c) (right): The same experimental phase results as shown in figure \ref{fig:fiveComparisons}, however the average of each experiment's measured values is set to be the same value - the value determined to be the mean of the offset distribution presented in.}
    
\label{fig:timeCompFigMasterLabel}

\end{figure}

The residual artifacts are easily seen in \Cref{fig:timeCompFigMasterLabel}, which shows the results from five chronological CO$_2$ measurements following the referencing feature described above.  
\Cref{fig:fiveComparisons} shows that the individual experiments reproduce similar trend lines across the sampled
photoelectron energies, however, the individual experiments still exhibit a systematic phase shift between experiments.
For each CO$_2$ dataset we again calculate the weighted sideband offset, Eqn.\ref{eqn:stageDriftMetric}, the distribution of these offsets is shown in \Cref{fig:phaseOffsetDistribution}.
Removing this weighted offset yields \Cref{fig:fiveComparisonsPostOffset}, which shows a greatly reduced variation in the measured phase offset. 

We interpret this behavior as a systematic uncertainty in the zero-phase measurement along with a random variation in the measured phase offset. 
Removing this systematic uncertainty using the weighted averaging technique produces a much smaller uncertainty in the variation of the retrieved phase.

\subsection{Comparative RABBITT}
\begin{figure}
\centering
\resizebox{0.45\columnwidth}{!}{\includegraphics{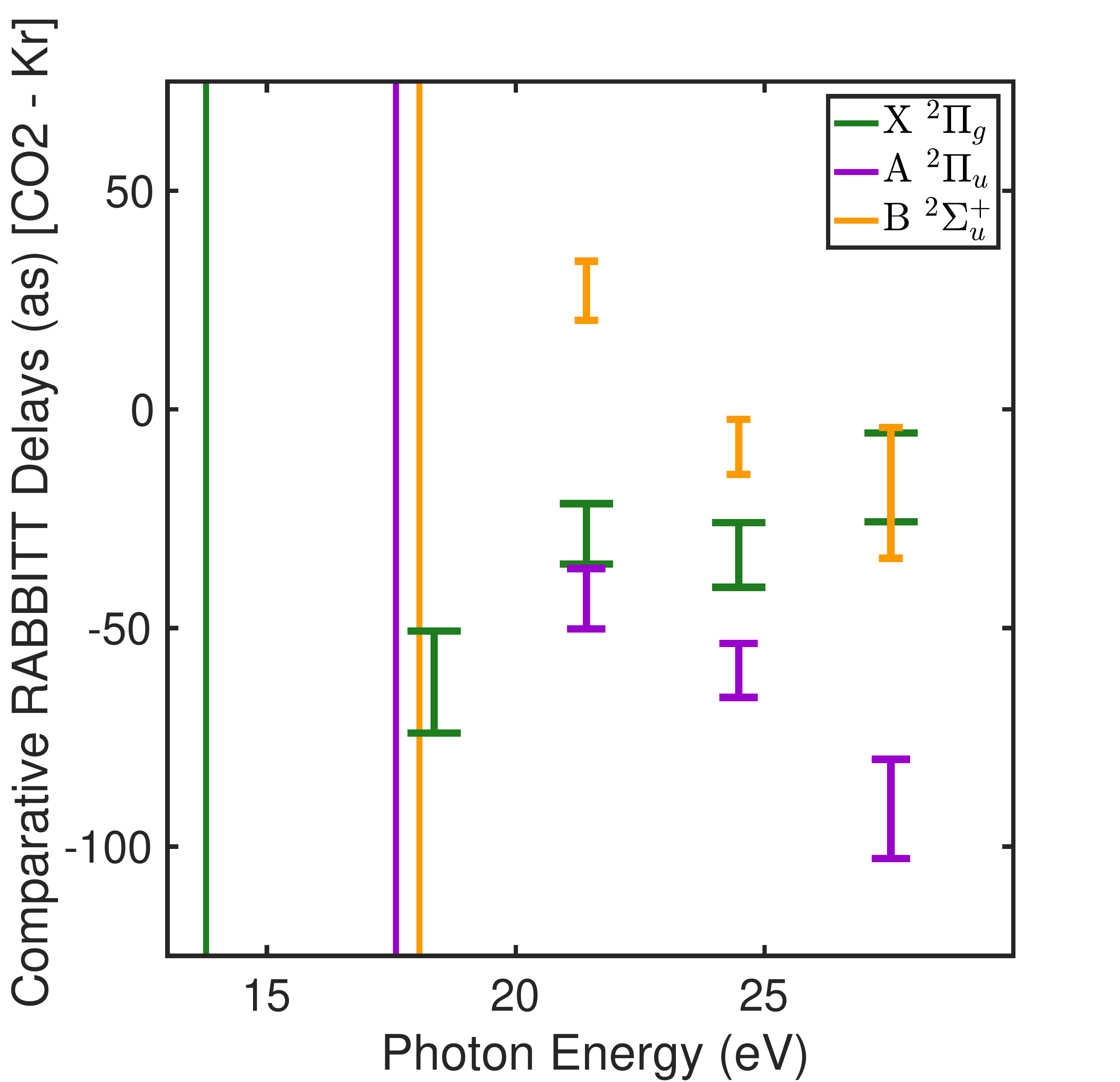}}
\caption{The comparative RABBITT findings for this work, as expressed in eqn. \ref{eqn:compRabEqnSupple}.  The vertical lines on the left side of the plot present the ionization thresholds for each of the three cationic states presented.  The errorbars represent $2\sigma$ uncertainty.  Not shown is a systematic error of $2\sigma = 26.2$~as associated with delay stage drift compensation.  This delay-stage error is a universal error value for all data points and can be thought of as an error in assigning the location of $\tau = 0$ on the axis.}
\label{fig:compRab}
\end{figure}
As described in the main paper, the extracted phase offset must be processed to isolate the phase component of interest, the photoionization phase, described in equations~2-5 of the main paper.
Referencing the measured time delay to a known target gas means that:
\begin{equation}
\label{eqn:compRabEqnSupple}
    \tau^{(I)}_{2}-\tau_2^{(\text{Kr})}=\Delta\tau^{(I)}_{\text{comp}}\approx\tau^{(I)}_{\text{PI}}-\tau^{\text{(Kr)}}_{\text{PI}}+\Delta\tau_{\text{c.c}}
\end{equation}
After applying the drift compensation procedure outlined above, we reference the extracted CO$_2$ two-photon photoionization time delays to those measured in krypton.
The referenced time-delays~($\Delta\tau^{(I)}_{\text{comp}}$) are shown in Figure~\ref{fig:compRab}.
To produce the data shown in Fig~1, panels~(a), we add the theoretical value of the photoionization time delay in krypton\cite{Kheifets2016RelativisticDelay} to the $X$-state measurements.
For panel~(b), we take the difference between the $A$ and $B$-state delays.
This is done to minimize the separation of ionization thresholds for the signals being compared.
Existing models for calculating $\Delta\tau_{\text{c.c}}$ lose accuracy at low photoelectron energies.
The inaccuracy can be sidestepped by minimizing the change in energy across which $\Delta\tau_{\text{c.c}}$ is calculated.
Results presented in the main paper use the $(P+A)'$ model \cite{dahlstrom_theory_2013}.
The $B - A$ data point near photon energy $\hbar\omega = 21.4$~eV was dropped because of $\Delta\tau_{\text{c.c}}$ model uncertainty.

\section{Theory}
\subsection{Channel Coupling in Two-Color Photoionization\label{subsect:twoPhotonDerivation}}
The interferometric ionization technique used in this work measures the delay in two-color photoionization defined as~\cite{dahlstrom_theory_2013,baykusheva_theory_2017}:
\begin{equation}
    \tau_n(2q, \hat{k}, \hat{R}) = \frac{1}{2\omega}\text{Arg}\left[M_n^{(2)*}(\vec{k};\epsilon_0+\Omega_{2q+1},\hat{R})M_n^{(2)}(\vec{k};\epsilon_0+\Omega_{2q-1},\hat{R})\right],
    \label{Eqn:Supp:TwoPhotonDelay}
\end{equation}
where, $\vec{k}$ is the final momentum of the outgoing electron for ionic channel $n$, $\hat{R}$ defines the orientation of the molecular axis relative to the laser polarization, and $\hbar\Omega_{2q\pm1}$ are the energies of the XUV photons which interfere to create the 2q$^{\text{th}}$ sideband. 
We assume that $\hbar\Omega_{2q\pm1}$ is larger than any of the relevant channel binding energies.
The two-photon ionization matrix element ($M_n^{(2)}$) for channel $n$ is given by:
\begin{equation}
    M^{(2)}_n(\vec{k}; \epsilon_0+\Omega,\hat{R})=-i E_{\omega}E_{X}\sum_p\int d\epsilon_i \frac{\langle \Psi^{(-)}_{\vec{k},n}\vert \vec{\mu}_{\omega}\cdot\vec{r}\vert \Psi_{i,p}\rangle \langle \Psi_{i,p}\vert \vec{\mu}_{\text{X}}\cdot\vec{r} \vert \Psi_0\rangle}{\epsilon_0+\hbar\Omega-\epsilon_i},
    \label{Eqn:Supp:TPME}
\end{equation}
where $\Psi^{(-)}_{\vec{k},n}$ is the observed final state wavefunction of momentum $\vec{k}$ in ionic channel $n$, and $\Psi_{i,p}$ is the intermediate state of energy $\epsilon_i$ in ionic channel $p$ produced by XUV ionization of the ground state, $\Psi_0$ with negative energy $\epsilon_0$. 
In Eqn.~\ref{Eqn:Supp:TPME} we have ignored the far off-resonance matrix element, where the IR pulse interacts before XUV pulse. 
Thus, Eqn.~\ref{Eqn:Supp:TwoPhotonDelay} is both emission angle- and orientation-resolved. 
Recently Douguet~\textit{et al.} developed a full numerical theory to describe $M^{(2)}_n$ using the complex Kohn variational method~\cite{douguet_application_2018}.
Here we develop an intermediate theory, similar to that originally developed by Dahlstrom~\textit{et al.}~\cite{dahlstrom_introduction_2012}, which focuses on the long-range behavior of the wavefunction. 
% in the complex Kohn approach. 
For simplicity, we assume that both the XUV and IR fields are co-polarized and linear.
Our approach for analyzing Eqn.~\ref{Eqn:Supp:TPME} is to express the scattering wavefunctions using a close coupling expansion
% expand the relevant matrix elements in the complex Kohn framework:
%
% \begin{eqnarray}
% |\Psi^{(-)}_{\vec{k},n}\rangle &=& \sum_p
%\left|\hat{A}\left[\Phi_p(\vec{x}_1,...,\vec{x}_n,\sigma_{N+1})
% F_{p,n\vec{k}}(\vec{r}_{N+1})\right]\right\rangle
% \label{Eqn:Supp:Kohn1}
% \end{eqnarray}
%where $\vec{x}_i=(\vec{r}_i,\sigma_i)$ are the position and spin variables of the $i^{\text{th}}$electron, $\Phi_p$ are the N-electron wavefunctions for the residual ion in channel $p$, and $F_{p,n\vec{k}}$ is a single electron scattering wavefunction.
% %The operator $\hat{A}$ ensures that the wavefuction is antisymmetric in the exchange of electrons. 
%The operator $\hat{A}$ antisymmetrizes the product $\Phi_p F_{p,n\vec{k}}$ 
% in the case of electron excahnge.
%
\begin{eqnarray}
\vert\Psi^{(-)}_{\vec{k},n}\rangle &=& \sum_p 
\int d^3\vec{r} \ F_{p,n\vec{k}}(\vec{r}) a^\dag(\vec{r})
\vert \Phi_p \rangle
\label{Eqn:Supp:Kohn1}
\end{eqnarray}
in which $\Phi_p$ are the (N-1)-electron wavefunctions for the residual 
ion in channel $p$, $a^\dag(\vec{r})$ is the creation
operator for a (properly spin-coupled) electron at position $\vec{r}$,
and $F_{p,n\vec{k}}$ is a single electron wavefunction.
The sum over cation channels $p$ in Eqn.~\ref{Eqn:Supp:Kohn1} is complete,
but the asymptotic form of the wavefunction concerns only the channels $p,n$
for which photoionization is allowed.
% - already mentioned above -
% In the present theory following B&W 
% we have considered only the long range behavior of the wavefunction
% in order to produce an analytic theory.
% We note that there are fully \textit{ab initio} methods 
% %whereas neglected terms describing short-range effects 
% in which the intermediate states are calculated % explicitly~\cite{douguet_application_2018}.
%
%
% The scattering orbitals are expanded in a set of continuum functions,
The asymptotic form of the single-electron wavefunctions $F_{p,n}$ 
for open channels is
encoded by the S-matrix,
\begin{equation}
    F_{p,n\vec{k}}(\vec{r}) \xrightarrow{r \rightarrow \infty} \sum_{\substack{LM\\lm}} Y_l^{m}(\hat{k}) \left[h^{(+)}_{\substack{nk\\lm}}(\vec{r})\delta_{\substack{np\\LM\\lm}}+h^{(-)}_{\substack{pk\\lm}}(\vec{r})S^*_{\substack{np\\LM\\lm}}(k) \right]
    \label{Eqn:Supp:Kohn2}
\end{equation}
where $Y_L^M$ are the spherical harmonic functions, and we have defined the regular-incoming~($h^{(+)}$) and irregular-outgoing~($h^{(-)}$) Coulomb waves as,
\begin{eqnarray}
\begin{split}
    h^{(+)}_{\substack{nk\\LM}} &=& \frac{1}{\sqrt{2}}\left(f_{\substack{nk\\LM}} + ig_{\substack{nk\\LM}}\right)\\
    h^{(-)}_{\substack{nk\\LM}} &=& \frac{1}{\sqrt{2}}\left(f_{\substack{nk\\LM}} - ig_{\substack{nk\\LM}}\right)
\end{split}
\quad   \quad
\begin{split}
    f_{\substack{nk\\LM}} &\xrightarrow{r\rightarrow\infty}&\frac{N_k}{r}\sin\left[kr+\Theta_{kL}(r)\right]Y_L^{M}(\hat{r})\\
    g_{\substack{nk\\LM}} &\xrightarrow{r\rightarrow\infty}&\frac{N_k}{r}\cos\left[kr+\Theta_{kL}(r)\right]Y_L^{M}(\hat{r}).
\end{split}
\end{eqnarray}
$f$~and~$g$ are the regular and irregular Coulomb function respectively, and $N_k=\sqrt{\nicefrac{2}{\pi k}}$ is a normalization constant. 
The phase of the Coulomb functions is given by,
\begin{equation}
    \Theta_{kL}(r) = \frac{1}{k}\text{ln}(2kr)-\frac{L\pi}{2}+\sigma_L,
\end{equation}
where $\sigma_L=\text{arg}\left[\Gamma(L+1-\nicefrac{i}{k})\right]$ is the scattering phase shift induced by the Coulomb-potential, and $\Gamma$ is the complex gamma function. 

Using Eqns.~\ref{Eqn:Supp:Kohn1}~and~\ref{Eqn:Supp:Kohn2} we can rewrite the matrix element between the intermediate and final states, $\langle\Psi^{(-)}_{\vec{k},n}\vert\vec{r}\vert \Psi_{i,p}\rangle$,
\begin{eqnarray}
    \langle\Psi^{(-)}_{\vec{k},n}\vert\vec{r}\vert \Psi_{i,p}\rangle &=& \sum_{qq^{\prime}}\left\langle \Phi_q F_{q,nk} \vert \vec{r} \vert \Phi_{q^{\prime}} F_{q^{\prime},p\kappa}\right\rangle \nonumber \\
    &=& \sum_{qq^{\prime}} \left\langle \Phi_q \vert \vec{r} \vert \Phi_{q^{\prime}}\right\rangle \left\langle F_{q,nk} \vert  F_{q^{\prime},p\kappa}\right\rangle + \delta_{qq^{\prime}}\left\langle F_{q,nk} \vert \vec{r} \vert  F_{q^{\prime},p\kappa}\right\rangle  \\
    &=& \sum_q \left\langle F_{q,nk} \vert \vec{r} \vert  F_{q,p\kappa}\right\rangle, \nonumber
\end{eqnarray}
where we have assumed that there are no resonant transitions between the residual ionic states.
Following the approach laid out by Dahlstrom~\textit{et al.} we rewrite Eqn.~\ref{Eqn:Supp:TPME} using Eqn.~\ref{Eqn:Supp:Kohn1},
\begin{equation}
    M^{(2)}_n(\vec{k}; \epsilon_0+\Omega,\hat{R}) = \frac{1}{i}\sqrt{\frac{4\pi}{3}}E_{\omega}E_X D^{(1)}_{00}(\hat{R})D^{(1)}_{00}(\hat{R})\sum_{q}\left\langle F_{q,nk}\vert rY_1^0(\hat{r})\vert\rho_{q,\kappa}\right\rangle,
\end{equation}
where $D^{(j)}_{lk}(\hat{R})$ is the Wigner matrix which rotates the polarization of the XUV and IR pulses into the molecular frame, and we have defined the first-order perturbed wavefunction for the q$^{\text{th}}$ channel, $\rho_{q,\kappa}(r)$ as
\begin{equation}
    \rho_{q,\kappa} = \sum_p
    \int d\epsilon_\kappa d\hat{\Omega}_{\hat{\kappa}} \frac{F_{q,p\kappa}(\vec{r}) \left\langle\left.\left.\psi^{(-)}_{\vec{\kappa},p}\right\vert z \right\vert\Phi_0\right\rangle}{\epsilon_0+\hbar\Omega-\epsilon_\kappa}\nonumber
\end{equation}
Using Eqn.~\ref{Eqn:Supp:Kohn2} as an equality we rewrite $\rho_{q,\kappa}$,
\begin{eqnarray}
\rho_{q,\kappa} &=& \sum_p\sum_{\substack{LM\\lm}} \int d\hat{\Omega}_{\hat{\kappa}}Y_L^M(\hat{\kappa})\left[\int d\epsilon_\kappa \delta_{\substack{qp\\LM\\lm}} \frac{h^{(+)}_{\substack{p\kappa\\lm}}(\vec{r}) \left\langle\psi^{(-)}_{\vec{\kappa},p}\vert z \vert\Phi_0\right\rangle}{\epsilon_0+\hbar\Omega-\epsilon_\kappa}\right.\\
&&+ \left.\int d\epsilon_\kappa S^*_{\substack{qp\\LM\\lm}}(\kappa)\frac{h^{(-)}_{\substack{p\kappa\\lm}}(\vec{r})\left\langle\psi^{(-)}_{\vec{\kappa},p}\vert z \vert\Phi_0\right\rangle}{\epsilon_0+\hbar\Omega-\epsilon_\kappa}\right]\nonumber
\end{eqnarray}
We rewrite the asymptotic form on Eqn.~12 using the following relations from Ref.~\cite{dahlstrom_theory_2013}
\begin{eqnarray}
\int d\epsilon_\kappa\frac{f_{\substack{p\kappa\\LM}} \left\langle\psi^{(-)}_{\vec{\kappa},p}\vert z \vert\Phi_0\right\rangle}{\epsilon_0+\hbar\Omega-\epsilon_\kappa} &=& \left.\left.-\frac{\pi N_\kappa }{r}Y_L^M(\hat{r})\exp\right[i\kappa r+i\Theta_{\kappa L}(r)\right]\left\langle\psi^{(-)}_{\vec{\kappa},p}\vert z \vert\Phi_0\right\rangle \nonumber \\
\int d\epsilon_\kappa\frac{g_{\substack{p\kappa\\LM}} \left\langle\psi^{(-)}_{\vec{\kappa},p}\vert z \vert\Phi_0\right\rangle}{\epsilon_0+\hbar\Omega-\epsilon_\kappa} &=& -i\left.\left.\frac{\pi N_\kappa }{r}Y_L^M(\hat{r})\exp\right[i\kappa r+i\Theta_{\kappa L}(r)\right]\left\langle\psi^{(-)}_{\vec{\kappa},p}\vert z \vert\Phi_0\right\rangle \nonumber
\end{eqnarray}
Then $\rho_{q,\kappa}$ becomes
\begin{eqnarray}
\rho_{q,\kappa} &=& -\frac{\pi N_\kappa}{r}\sum_p\sum_{\substack{LM\\lm}}Y_l^m(\hat{r}) S_{\substack{pq\\LM\\lm}}^*(\kappa)\exp[i\kappa r+i\Theta_{\kappa l}(r)] 
I_{p,lm}(\kappa)
%\left[\sum_{q^{\prime}}\left\langle\Phi_{q^{\prime}}F_{q^{\prime},p\kappa}\vert r \vert\Phi_0\right\rangle\right]_{lm}
\end{eqnarray}
where
\begin{equation}
    %\left[\sum_{q^{\prime}}\left\langle\Phi_{q^{\prime}}F_{q^{\prime},p\kappa}\vert r \vert\Phi_0\right\rangle\right]_{lm} 
    I_{p,lm}(\kappa) = \int d\hat{\Omega}_{\hat{\kappa}}Y_l^m\left(\hat{\kappa}\right)\left.\left.\left\langle\psi^{(-)}_{\vec{\kappa},p}\right\vert z \right\vert\Phi_0\right\rangle
    \label{Eqn:Supp:OnePhoton}
\end{equation}
is the partial wave expansion of the XUV dipole matrix element for ionization to the $p$ channel, which is calculated using the complex Kohn approach, and described in the next section.

Next we must evaluate the matrix element, $\left\langle F_{q,nk}\vert r\vert\rho_{q,p\kappa}\right\rangle$
\begin{eqnarray}
    \left\langle F_{q,nk}\vert r Y_1^0\vert\rho_{q\kappa}\right\rangle &=& -\pi N_k N_\kappa \sum_p\sum_{\substack{LM\\L^\prime M^\prime}}Y_L^M(\hat{k})\sum_{\substack{lm\\l^\prime m^\prime}}\left\langle Y_l^m\vert Y_1^0\vert Y_{l^\prime}^{m^\prime}\right\rangle I_{p,l^\prime m^\prime}(\kappa) \nonumber \\
    &&\times \left[S^*_{\substack{pq\\L^\prime M^\prime\\l^\prime m^\prime}}(\kappa)\delta_{\substack{nq\\LM\\lm}}\int_0^\infty dr \exp\left[i(\kappa-k)r\right]r\exp\left[i(\Theta_{\kappa}(r)-\Theta_k(r))\right]\right. \\
    &&\left.+ S^*_{\substack{pq\\L^\prime M^\prime\\l^\prime m^\prime}}(\kappa) S_{\substack{qn\\LM\\lm}}(k)\int_0^\infty dr \exp\left[i(\kappa+k)r\right]r\exp\left[i(\Theta_{\kappa}(r)+\Theta_k(r))\right] \right] \nonumber
\end{eqnarray}
We define the following integrals, 
\begin{equation}
    J_{\pm}(\kappa,k)= \pm\frac{1}{2i}\int_0^\infty dr~r^{1+i(\nicefrac{1}{\kappa}\pm\nicefrac{1}{k})}\exp[i(\kappa\pm k)r] \nonumber
\end{equation}
and rewrite Eqn.~15,
\begin{eqnarray}
    \left\langle F_{q,nk}\vert r\vert\rho_{q\kappa}\right\rangle &\sim& -\pi N_\kappa N_k \sum_p\sum_{\substack{LM\\L^\prime M^\prime}}Y_L^M(\hat{k})\sum_{\substack{lm\\l^\prime m^\prime}}\left\langle Y_l^m\vert Y_1^0\vert Y_{l^\prime}^{m^\prime}\right\rangle I_{p,l^\prime m^\prime}(\kappa) \\
    &&\times S^*_{\substack{pq\\L^\prime M^\prime\\l^\prime m^\prime}}(\kappa) \left(\delta_{\substack{nq\\LM\\lm}}J_{-} + S_{\substack{qn\\LM\\lm}}(k) J_{+}\right) \nonumber
\end{eqnarray}
As described in Ref.~\cite{dahlstrom_theory_2013}, the contribution from the $J_+$ integral is much smaller than that of the $J_{-}$ integral. 
This is because of the IR photon energy is small compared to the final electron energy, and thus $\nicefrac{k^2}{2}$-$\nicefrac{\kappa^2}{2}=\omega<\nicefrac{k^2}{2}$. 
Therefore the difference $\vert \kappa-k \vert$ is much smaller than the sum $\kappa+k$ and the fast oscillations of $\exp[i(\kappa+k)r]$ in the $J_{+}$ integral lead to a cancellation.
Thus we arrive at the expression,
\begin{eqnarray}
    \left\langle F_{q,nk}\vert r\vert\rho_{q\kappa}\right\rangle &=& -\pi N_k N_\kappa \sum_p\sum_{\substack{LM\\L^\prime M^\prime\\l^\prime m^\prime}}Y_L^M(\hat{k})I_{p,l^\prime m^\prime}(\kappa)\left\langle Y_L^M\vert Y_1^0\vert Y_{l^\prime}^{m^\prime}\right\rangle S^*_{\substack{pq\\L^\prime M^\prime\\l^\prime m^\prime}}(\kappa) \delta_{\substack{nq}}J_{-}(\kappa,k).%\nonumber
\end{eqnarray}

Using Eqn.~17 we arrive at the following simplification for Eqn.~\ref{Eqn:Supp:TPME}
\begin{eqnarray}
M^{(2)}_n(\vec{k}; \epsilon_0+\Omega,\hat{R}) &=& \frac{4\pi^2}{3i} E_{\omega}E_X D^{(1)}_{00}(\hat{R})D^{(1)}_{00}(\hat{R})N_k N_\kappa \sum_{p}\sum_{\substack{LM\\L^\prime M^\prime}}Y_L^M(\hat{k}) \\
&&\times \sum_{\substack{l^\prime m^\prime}}I_{p,l^\prime m^\prime}(\kappa)\left\langle Y_L^M\vert Y_1^0\vert Y_{l^\prime}^{m^\prime}\right\rangle S^*_{\substack{pn\\L^\prime M^\prime\\l^\prime m^\prime}}(\kappa)J_{-}(\kappa,k).\nonumber
\end{eqnarray}
This can be recast in the same form as the expression derived by Baykusheva and Worner~\cite{baykusheva_theory_2017}:
\begin{equation}
  M^{(2)}_n(\vec{k}; \epsilon_0+\Omega,\hat{R}) = N_k N_\kappa J_{-}(k,\kappa) \sum_{LM} b_{n,LM}(\kappa; \hat{R}) Y_L^M(\hat{k})  
\end{equation}
But now $b_{n,LM}$ is defined as,
\begin{eqnarray}
    \label{Eqn:Supp:bLM}
    b_{n,LM}(\kappa; \hat{R}) &=& \frac{4\pi}{3i} E_{\omega}E_X D^{(1)}_{00}(\hat{R})D^{(1)}_{00}(\hat{R})\sum_{p}\sum_{\substack{L^\prime M^\prime\\l^\prime m^\prime}}I_{p,l^\prime m^\prime}(\kappa)\left\langle Y_L^M\vert Y_1^0\vert Y_{l^\prime}^{m^\prime}\right\rangle S^*_{\substack{pn\\L^\prime M^\prime\\l^\prime m^\prime}}(\kappa)%\\
    %I_{p,LM} &=& \left[ \sum_{q^{\prime}}\left\langle\Phi_{q^{\prime}}F_{q^{\prime},p\kappa}\vert r \vert\Phi_0\right\rangle\right]_{L^\prime M^\prime} \nonumber
\end{eqnarray}
The angle- and orientation-dependent delay in Eqn.~\ref{Eqn:Supp:TwoPhotonDelay} is then given by
\begin{eqnarray}
    \label{Eqn:Supp:TwoPhotonDelayResult}
    \tau_n(2q,\hat{k},\hat{R}) &=& \frac{1}{2\omega}\text{Arg}\left[N_{k-\omega} J_{-}(k,k-\omega) N^*_{k+\omega} J^*_{-}(k,k+\omega)\right]\\
    &&+ \frac{1}{2\omega}\text{Arg}\left[\sum_{\substack{LM\\L^\prime M^\prime}}Y_L^{M*}(\hat{k})Y_{L^\prime}^{M^{\prime}}(\hat{k})b^*_{n,LM}(k-\omega;\hat{R})b_{n,L^\prime M^\prime}(k+\omega;\hat{R})\right]\nonumber \\
    &=& \tau_{cc}(k) + \tau_{PI}(2q,\hat{k},\hat{R}) \nonumber
\end{eqnarray}
As pointed out in Ref.~\cite{baykusheva_theory_2017}, the first term in Eqn.~\ref{Eqn:Supp:TwoPhotonDelayResult} only depends on the photon energy and can be interpreted as a continuum-continuum delay~($\tau_{cc}$).
The second term in Eqn.~\ref{Eqn:Supp:TwoPhotonDelayResult} is again a target-specific delay~($\tau_{PI}$), but the single photon delays are modified by the interaction with the IR light.
In addition to this modification, described by Baykusheva and Worner, Eqn.~\ref{Eqn:Supp:bLM} shows that the channel coupling adds additional modifications. 
In the absence of any channel coupling, the $S$-matrix is purely diagonal, Eqn.~\ref{Eqn:Supp:bLM} reduces to the expression derived by Baykusheva and Worner.

%where we have defined the integrals $J_{\pm}$ as,
%\begin{equation}
%    J_{\pm}(\kappa,k)= \pm\frac{1}{2i}\left(\frac{i}{\kappa\pm k}\right)^{2+i(\nicefrac{1}{\kappa}\pm\nicefrac{1}{k})}\Gamma\left[2+i\left(\frac{1}{\kappa}\pm\frac{1}{k}\right)\right]\nonumber
%\end{equation}

\subsection{Complex Kohn Photoionization Calculations}
\begin{table}
\begin{tabular}{l | ccccc}
State & Ground neutral & $X \ ^2\Pi_g$ & $A \ ^2\Pi_u$ & $B \ ^2\Sigma_u$ & $C \ ^2\Sigma_g$ \\
Energy &	-187.654	& -187.150 &	-186.964 &	-186.947 &	-186.894 \\
IP	&	0.504 &	0.690 &	0.707	& 0.760 \\
IP (eV)	&	13.703 &	18.774	& 19.232	 & 20.677 \\
IP (lit.)   &   13.669 \\
   \end{tabular}
   \caption{Energies (in hartree or eV) of the ground 
and cation electronic states obtained from this calculation at R=2.140
compared with benchmark aug-pvtz coupled cluster results [Can't immediately find higher IPs.
NIST CCCBDB has CO2 vertical IP = 13.669eV using aug-cc-pVTZ, CCSD(T).  https://cccbdb.nist.gov/ie2x.asp?casno=124389 ]}
\label{ktab2}
\end{table}
In this section we describe how we calculate the single-photon, channel-resolved dipole matrix element~(Eqn.~11 of the main text), 
\begin{equation}
    \label{eqn:supp:dipoleMatrixElementKohn}
    d_{n}(\vec{k})=\sum_{LM}Y_L^M(\hat{k})I_{n,LM}(k),
\end{equation}
described in Eqn.~\ref{Eqn:Supp:OnePhoton}.
As described in the main text,
the channel-resolved dipole matrix element
%, Eqn.~\ref{Eqn::DME}, 
is computed using the complex Kohn method for photoionization~\cite{\citekohnimp,\citekohnphoto}, using
%The calculations use explicit 
representations of the initial neutral state and of the final cation states obtained with
one single 11-orbital basis.  The orbital basis 
% This basis for neutral and cation 
was obtained using 
%a state-averaged multiconfiguration self-consistent-field~(MCSCF) calculation performed with 
the \textit{Columbus} quantum chemistry program~\cite{Lischka_New_2009,Shepard_Progress_1988,Lischka_High_2001,Lischka_Columbus_2011,Coumbus7}.
The orbitals are obtained by minimizing 
a weighted sum of energies of the closed-shell Hartree-Fock ground state neutral and six microstates
of the cation, corresponding to 
the $X \ ^2\Pi_g$, $A \ ^2\Pi_u$, $B \ ^2\Sigma_u$, and $C \ ^2\Sigma_g$ electronic states, at 
bond distance $R=2.140a_0$.
The relative weights for ground and cation are 6:1:1:1:1:1:1, such that 
the orbitals obtained are a balanced compromise between those optimized for the neutral and cation.
The energies of these states are listed in Table~\ref{ktab2}.
The primitive basis for this MCSCF calculation was Dunning's \textit{aug-cc-pVDZ} basis set~\cite{Dunning_Gaussian_1989}, comprising 75 basis functions, 
with 35
% 25 
additional basis functions on the carbon atom (at the origin)
that are listed in Table~\ref{ktab1}.
For the complex Kohn scattering calculation, the $L=0,1,2,3$ partial wave scattering channels are used,
\begin{equation}
    \left\vert\Psi^{(-)}_{\vec{k},n}\right\rangle = \sum_{LM} Y_L^M\left(\hat{k}\right) \left\vert\psi^{(-)}_{kLM,n}\right\rangle
\end{equation}
Eleven linear combinations of the 110 Gaussian basis functions are obtained by
the state-averaged MCSCF calculation, and the orthogonal complement of 99 basis functions 
are used along with numerical continuum functions with $L=0,1,2,3$ to construct the scattering wave functions
$\psi^{(-)}_{kLM,n}$, which is described by Eqn.~\ref{Eqn:Supp:Kohn1}.
The dipole matrix element between the scattering wavefunction and the initial state wavefunction is calculated within the approximation of  separable exchange~\cite{Rescigno_Separable_1981,Rescigno_Disappearance_1988,McCurdy_Beyond_1992}.
Two different complex Kohn calculations are performed, one in which the four electronic state 
channels are uncoupled from one another, and one in which they are fully coupled, in the construction
of $\vert\psi^{(-)}\rangle$.
The dipole matrix element is expanded in a partial wave basis as described in Eqn.~\ref{eqn:supp:dipoleMatrixElementKohn}

\begin{table}
\begin{tabular}{r | cccc | l}
Partial wave & Exponents & & & & Number of functions \\
\hline
s    &  0.0200   & 0.0080   &      0.0030       &   0.0012      & 4 \\
 p   &  0.0180   & 0.0080    & 0.0035  & & 9 \\
 d   &  0.0750   & 0.0320  & & & 12 \\
 f    &  0.2500 & & & & 10 \\
 \hline
 all & & & & & 35 \\
\end{tabular}
\caption{Extra basis functions used in the Kohn calculation, those besides
Dunning's aug-cc-pVDZ basis~\cite{Dunning_Gaussian_1989}.}
\label{ktab1}
\end{table}

\subsection{Molecular Photoionization Delays}
\begin{figure}
\centering
    \begin{subfigure}{\columnwidth}
        \fcolorbox{black}{white}{
        \resizebox{0.98\columnwidth}{!}{\includegraphics{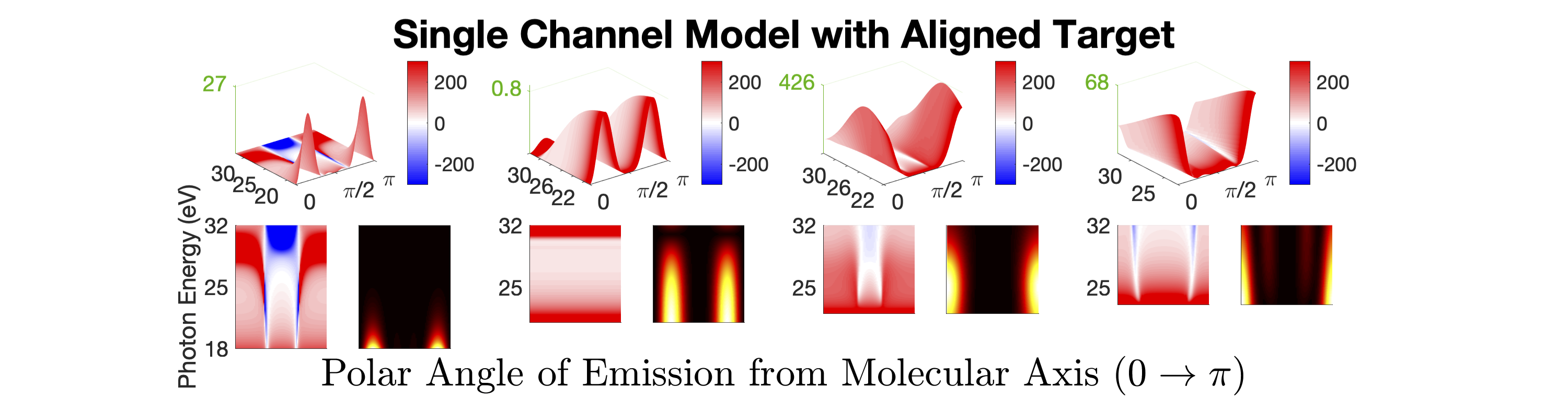}}}
        \phantomsubcaption
        \label{fig:supp:paraUncoup}
    \end{subfigure}
    \begin{subfigure}{\columnwidth}
        \fcolorbox{black}{white}{
        \resizebox{0.98\columnwidth}{!}{\includegraphics{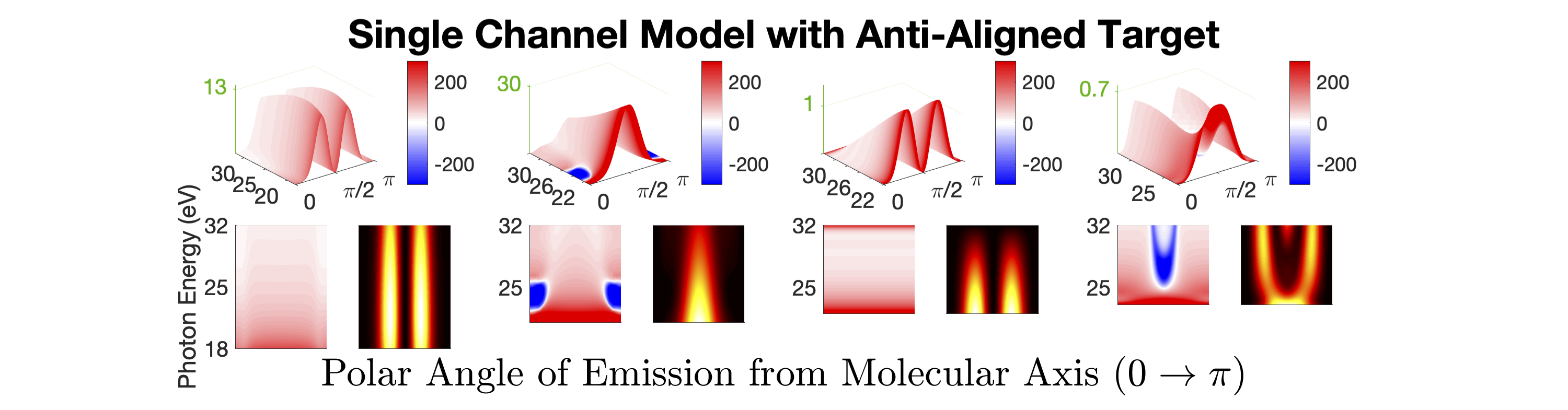}}}
        \phantomsubcaption
        \label{fig:supp:perpUncoup}
    \end{subfigure}
\caption{Calculated molecular frame two-photon photoionization time delays, $\tau_{PI}(2q,\hat{k},\hat{R})$, for different molecular orientations. In each box, the upper row of plots shows polar surfaces, as a function of photon energy and photoelectorn emission angle, where the topography is defined by the photoionization cross section and colormap by the photoioniaztion delay time. Below each contour plot, we show the calculated photoionization time delay on the left and calculated cross section on the right. Moving across each row, we show the results for ionization in the $X^2\Pi_g$, $A^2\Pi_u$, $B^2\Sigma_u$, and $C^2\Sigma_g$ cationic channels. The green number along the z-axis of the surface plot provides context for the relative cross sections of this particular channel. We show results for single channel calculations, where channel coupling is neglected, for the case when both the XUV and IR polarizations are linear and either aligned with the molecular axis~(Aligned target) or perpendicular to the molecular axis~(Anti-Aligned Target).  }
\label{fig:supp:SingleMFdelay}
\end{figure}

\begin{figure}
\centering
    \begin{subfigure}{\columnwidth}
        \fcolorbox{black}{white}{
        \resizebox{0.98\columnwidth}{!}{\includegraphics{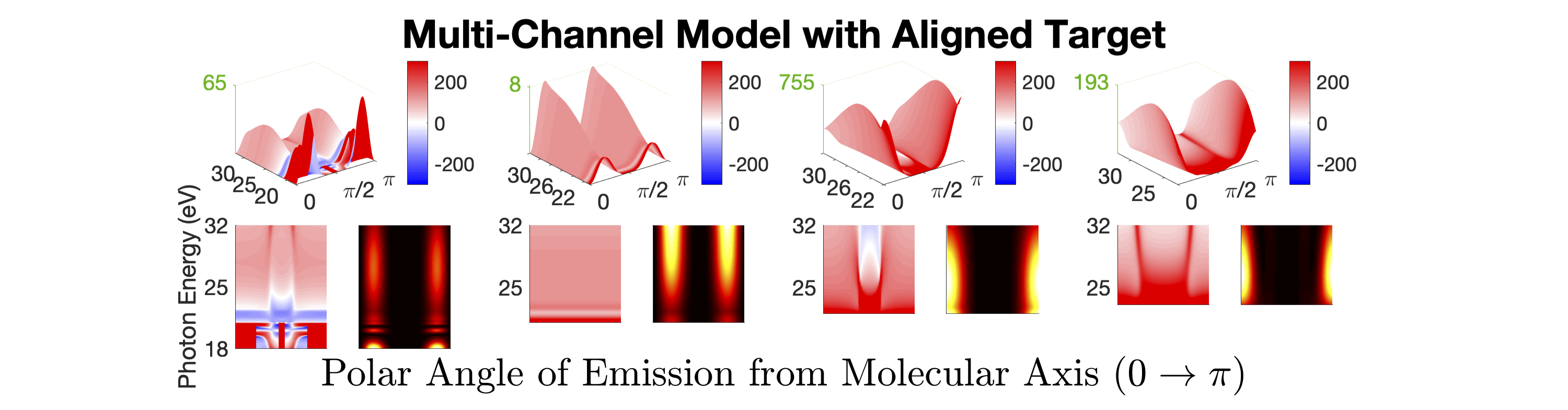}}}
        \phantomsubcaption
        \label{fig:supp:paraCoup}
    \end{subfigure}
    \begin{subfigure}{\columnwidth}
        \fcolorbox{black}{white}{
        \resizebox{0.98\columnwidth}{!}{\includegraphics{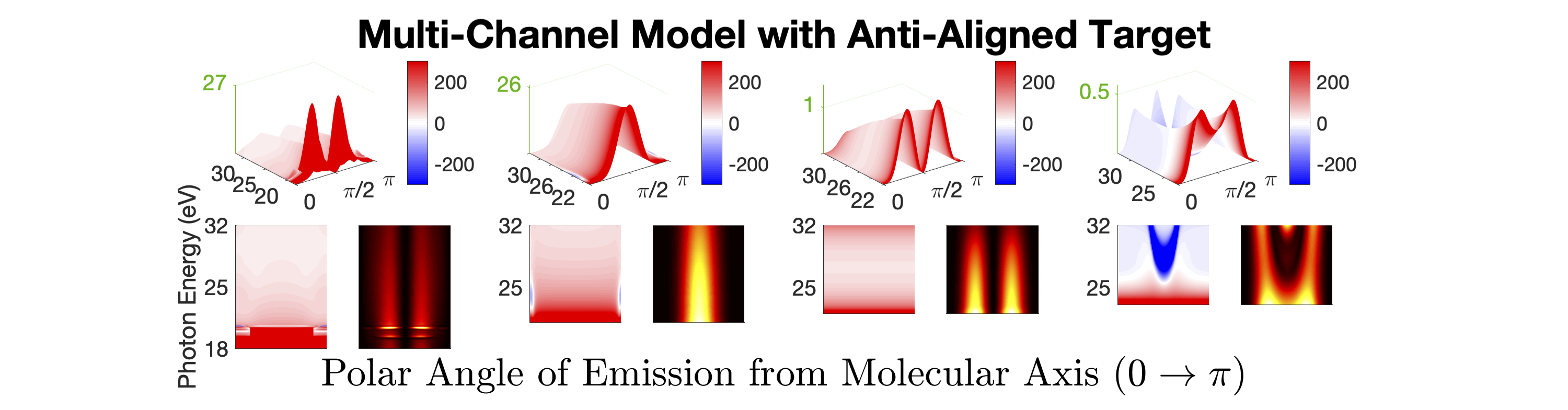}}}
        \phantomsubcaption
        \label{fig:supp:perpCoup}
    \end{subfigure}
    \caption{Calculated molecular frame two-photon photoionization time delays, $\tau_{PI}(2q,\hat{k},\hat{R})$, for different molecular orientations. In each box, the upper row of plots shows polar surfaces, as a function of photon energy and photoelectorn emission angle, where the topography is defined by the photoionization cross section and colormap by the photoioniaztion delay time. Below each contour plot, we show the calculated photoionization time delay on the left and calculated cross section on the right. Moving across each row, we show the results for ionization in the $X^2\Pi_g$, $A^2\Pi_u$, $B^2\Sigma_u$, and $C^2\Sigma_g$ cationic channels. The green number along the z-axis of the surface plot provides context for the relative cross sections of this particular channel. We show results for the coupled-channel calculation described above, for the case when both the XUV and IR polarizations are linear and either aligned with the molecular axis~(Aligned target) or perpendicular to the molecular axis~(Anti-Aligned Target).  }
\label{fig:supp:MultiMFdelay}
\end{figure}

%Need to begin with a description of how the coupling is done.
%Dipole moments $I_{p,lm}$ calculated with fully coupled continuum, whereas coupling in two-photon matrix element is neglected, i.e. $S$-matrix is diagonal in expression for $b_{lm}$
%For the single channel calculations this is exactly correct. 
The partial wave decomposition of the single-photon dipole matrix element, $I_{n,LM}(k)$ is calculated as described in the previous section. 
The results are used to calculate $b_{LM}(k;\hat{R})$ described in Eqn.~\ref{Eqn:Supp:bLM}.
For the calculation presented in the main text and below, we make the assumption that the $S$-matrix in Eqn.~\ref{Eqn:Supp:bLM} is purely diagonal.% i.e. $S^*_{\substack{pn\\L^\prime M^\prime\\l^\prime m^\prime}}(\kappa)=\delta_{\substack{pn\\L^\prime M^\prime\\l^\prime m^\prime}}$. 
In this approximation, we fully include channel coupling in the single-photon ionization process, but neglect the effect of the channel coupling in the interaction with the IR-photon.  
%Combining the Complex Kohn results of eqn. \ref{eqn:supp:dipoleMatrixElementKohn} with the derivation of eqn. \ref{Eqn:Supp:bLM}, we can solve for $\tau_{PI}(2q,\hat{k},\hat{R})$ in eqn. \ref{Eqn:Supp:TwoPhotonDelayResult}.
The result is the emission angle~($\hat{k}$)- and orientation angle~($\hat{R}$)-resolved photoionization delays for a central photon energy of $\hbar\Omega_{2q}$.% when measured with a XUV+IR photoionization setup, like the one used in this work.
The molecular frame~(MF) photoionization delays are shown in figures~\ref{fig:supp:SingleMFdelay}~and~\ref{fig:supp:MultiMFdelay}.
The delays are calculated in the two limiting cases where the XUV~(and IR) polarization is aligned along the molecular axis~(\Cref{fig:supp:paraUncoup,fig:supp:paraCoup}), and when this polarization is perpendicular to the molecualr axis~(\Cref{fig:supp:perpUncoup,fig:supp:perpCoup}).
We show the results of both the single channel~(\Cref{fig:supp:SingleMFdelay}) and coupled-channel~(\Cref{fig:supp:MultiMFdelay}) calculations.

%$\tau_{PI}(2q,\hat{k},\hat{R})$ for the (left to right) $X \ ^2\Pi_g$, $A \ ^2\Pi_u$, $B \ ^2\Sigma_u$, and $C \ ^2\Sigma_g$ cationic states for the case of XUV+IR parallel to the molecular axis for calculations that do not include channel coupling.  For each state panel, the x-axis is the polar angle of emission and the y-axis is the sideband photon energy $2q$.  The surface height corresponds to channel's cross section in arbitrary units (bottom right of each panel, unique color map) and the color corresponds to the photoionization delay (as)(bottom left of every panel).  The green number along the z-axis provides context for relative cross sections amongst the channels.  The jacobian term for the azimuthal angle of emission has not been weighed in; the plots represent delay values for any unspecified azimuthal angle.

\subsection{Effects of Angular Averaging}
%Eqn:Supp:tauPIDefinition is equation for tauPI
%Eqn:Supp:bLM is equation for b_lm terms that have the angle changing operators built in
There are a number of interesting features in the MF, angle-resolved photoemission delays, however, our experiments use a randomly oriented target gas and magnetic bottle electron time of flight spectrometer which does not resolve emission angle.
Therefore Eqn.~\ref{Eqn:Supp:TwoPhotonDelayResult} needs to be averaged over both the molecular axis distribution or the outgoing electron angle.
The contributions from each pair of $(\hat{k},\hat{R})$ are incoherently averaged, weighed by a cross section term $\vert b^*_{n,LM}(k-\omega;\hat{R})b_{n,L^\prime M^\prime}(k+\omega;\hat{R})\vert $.
The measured photoionization time delays are then compared to the calculated average, \begin{equation}
    \tau_{\text{PI}}(2q) = \frac{1}{2\omega}\text{Arg}\left[\int d\hat{\Omega}_{\hat{R}}\int d\hat{\Omega}_{\hat{k}} \sum_{\substack{LM\\L^\prime M^\prime}}Y_L^{M*}(\hat{k})Y_{L^\prime}^{M^{\prime}}(\hat{k})b^*_{n,LM}(k-\omega;\hat{R})b_{n,L^\prime M^\prime}(k+\omega;\hat{R})\right].
    \label{eqn:supp:twoPhotonAngleAveragedDelay}
\end{equation}

\begin{figure}
\centering
\begin{subfigure}{0.49\columnwidth}
        \resizebox{0.99\columnwidth}{!}{\includegraphics{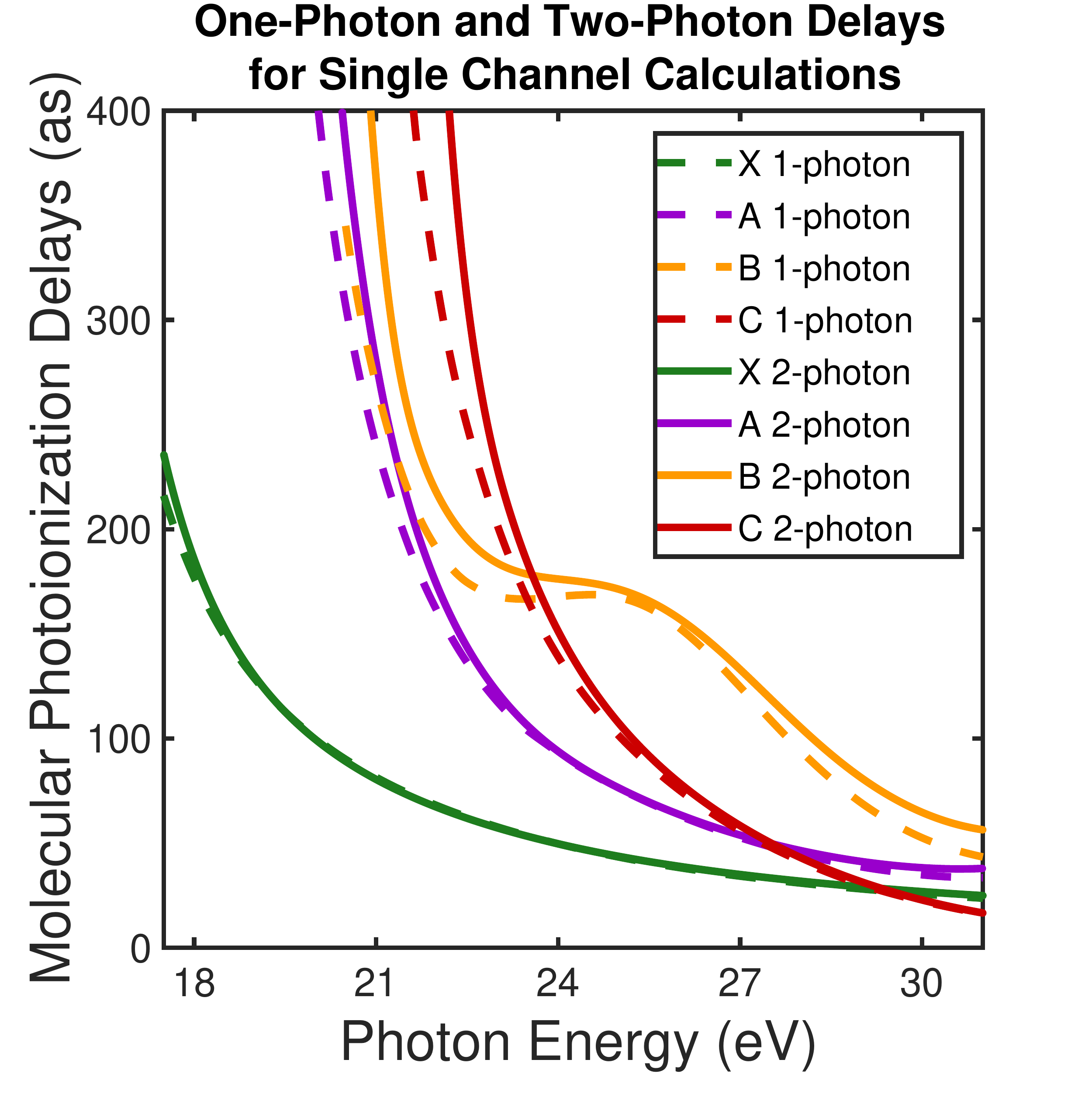}}
        \phantomsubcaption
        \label{fig:supp:onePhotonComparedtoTwoPhoton_uncoupled}
    \end{subfigure}
    \begin{subfigure}{0.49\columnwidth}
        \resizebox{0.99\columnwidth}{!}{\includegraphics{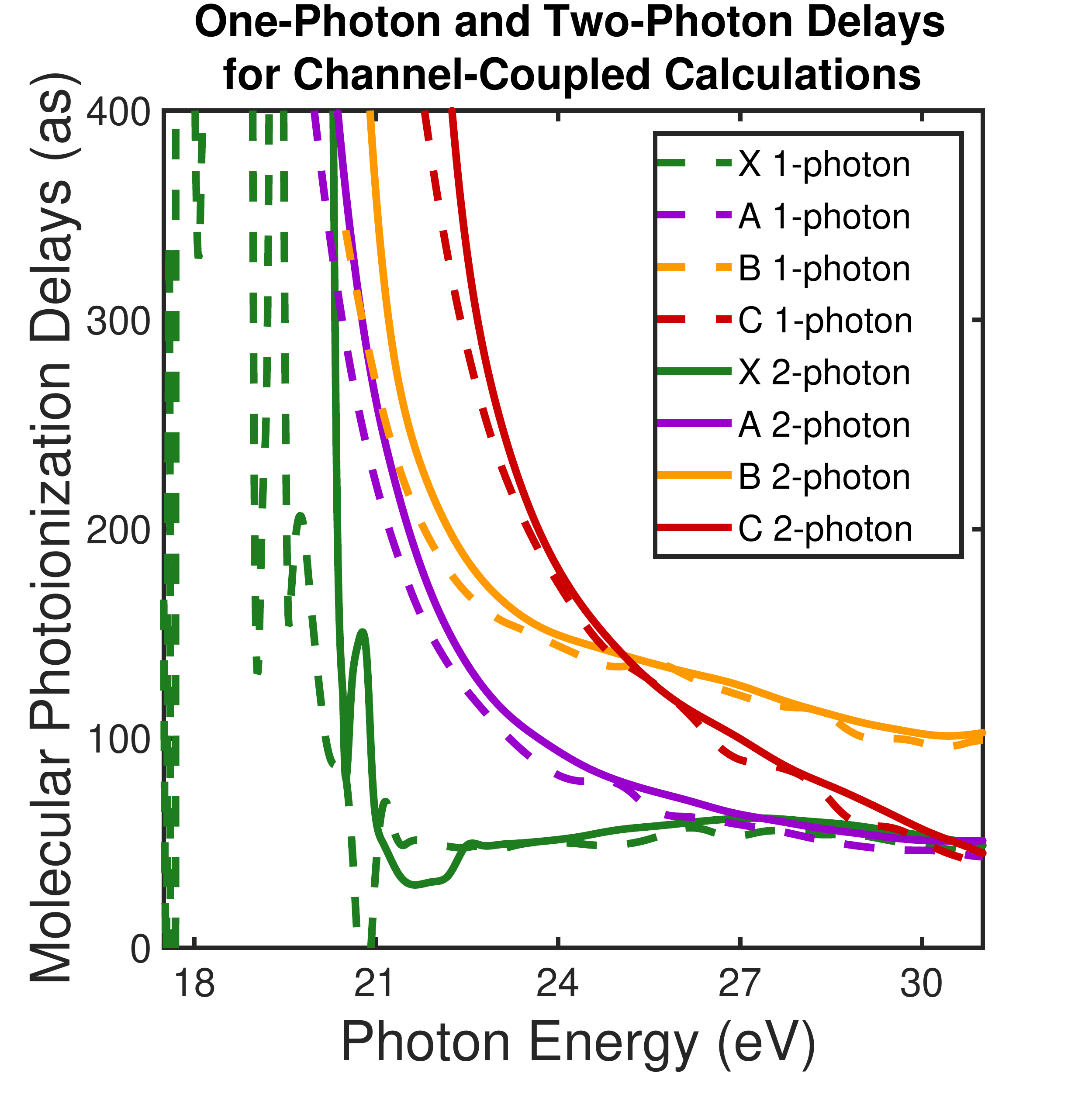}}
        \phantomsubcaption
        \label{fig:supp:onePhotonComparedtoTwoPhoton_coupled}
    \end{subfigure}
    \caption{Comparison of angle-averaged single-photon and angle-averaged two-photon photoionization delay values for the single channel~(left) and coupled-channel~(right) results.  Most features are consistent between $\tau_{1}$ and $\tau_{P.I.}$.  There is difference in the low-energy region for the X state results.  The Rydberg series consists of very long delays at discrete energies for the single-photon values, which get smeared into a plateau for the two-photon calculation.}
    \label{fig:supp:onePhotonComparedtoTwoPhoton}
\end{figure}

Eqn.~\ref{eqn:supp:twoPhotonAngleAveragedDelay} is an averaged quantity that is compared to the experimental measurements in Figure~1 of the main text. 
We can preform similar averaging for the single-photon photoionzation delay, $\tau_1(E,\hat{k},\hat{R})$, defined in Eqn.~9 of the main text. 
In \Cref{fig:supp:onePhotonComparedtoTwoPhoton} we compare the emission angle- and orientation-averaged single-photon~(($\tau_{1}(E)$) and two-photon~($\tau_{\text{PI}}(2q)$) delay, and show that the two values are nearly identical.
This is in stark contrast to the previous work by Baykusheva and Worner, where it was observed that for the N$_2$O target, there were substantial difference between the two values~\cite{baykusheva_theory_2017}. 
This is likely due to difference in the molecular system, such as the presence of inversion symmetry in the CO$_2$ system, which should simplify the partial wave expansion of the photoionized EWP, and lessen the effects that IR induced angular momentum coupling has on the emission angle-averaged photoionizaion delay times. 
%Similar effect were observed by Baykusheva and Worner in N$_2$~\cite{baykusheva_private}.

%In personal communications with D.~Baykusheva, we discussed that her preliminary work on N$_2$ also showed only minor discrepancy between single-photon and two-photon models.
%We speculate that this is likely due to difference in the molecular system, such as the presence of inversion symmetry which could simplify the partial wave expansion for the photoionized EWP and lessen the effects that IR induced angular momentum coupling has on observable photoionizaion delay times.

%\bibliography{referencesJPC.bib,co2_2019_djh_kohn.bib,co2_2019_djh_other.bib}

%apsrev4-2.bst 2018-12-27 (MD) hand-edited version of apsrev4-1.bst